\documentclass[prd,twocolumn,nofootinbib,superscriptaddress,fleqn,amssymb,amsfonts,amsmath]{revtex4-1}
\usepackage{graphicx,slashed,xcolor,multirow,bbold,mathtools,sidecap,tikz,bm,ulem,enumitem,booktabs,array}
\usepackage[left=0.6in,right=0.4in,top=0.5in,bottom=0.5in]{geometry}
\usepackage[colorlinks=true,linktoc=page,linkcolor=purple,citecolor=teal,urlcolor=magenta]{hyperref}
\usepackage[tiny,raggedright]{titlesec} 
\graphicspath{ {figures/} }

\allowdisplaybreaks
\setenumerate{label=$(\roman*)$,itemsep=1pt,parsep=1pt,topsep=1pt,partopsep=1pt}
\let\emph\textit

\AtBeginDocument{%
  \heavyrulewidth=.08em
  \lightrulewidth=.05em
  \cmidrulewidth=.03em
  \belowrulesep=.65ex
  \belowbottomsep=0pt
  \aboverulesep=.4ex
  \abovetopsep=0pt
  \cmidrulesep=\doublerulesep
  \cmidrulekern=.5em
  \defaultaddspace=.5em
}

\AtBeginDocument{%
\newwrite\bibnotes
\def\bibnotesext{Notes.bib}
\immediate\openout\bibnotes=\jobname\bibnotesext
\immediate\write\bibnotes{@CONTROL{REVTEX41Control}}
\immediate\write\bibnotes{@CONTROL{apsrev41Control,author="08",editor="1",pages="1",title="0",year="1"}}
\if@filesw
\immediate\write\@auxout{\string\citation{apsrev41Control}}%
\fi}

\definecolor{lime}{HTML}{A6CE39}
\DeclareRobustCommand{\orcidicon}{\hspace{-2.1mm}
\begin{tikzpicture}
\draw[lime,fill=lime] (0,0.0) circle [radius=0.13] node[white] {{\fontfamily{qag}\selectfont \tiny ID}}; \draw[white,fill=white] (-0.0525,0.095) circle [radius=0.007]; 
\end{tikzpicture} \hspace{-3.7mm} }
\foreach \x in {A, ..., Z}
 {\expandafter\xdef\csname orcid\x\endcsname{\noexpand\href{https://orcid.org/\csname orcidauthor\x\endcsname} {\noexpand\orcidicon}}}

\begin{document}

\title{Type-II see-saw: searching the LHC elusive low-mass triplet-like Higgses at $e^-e^+$ colliders}

\author{Saiyad Ashanujjaman\orcidA{}}
\email{saiyad.a@iopb.res.in}
\author{Kirtiman Ghosh}
\email{kirti.gh@gmail.com}
\affiliation{Institute of Physics, Bhubaneswar, Sachivalaya Marg, Sainik School, Bhubaneswar 751005, India}                                                           
\affiliation{Homi Bhabha National Institute, Training School Complex, Anushakti Nagar, Mumbai 400094, India}
\author{Katri Huitu}
\email{katri.huitu@helsinki.fi}
\affiliation{Department of Physics, and Helsinki Institute of Physics, University of Helsinki, Finland 00014}
                                                                                                                                                                                      
\begin{abstract}
\noindent While the triplet-like Higgses up to a few hundred GeV masses are already excluded for a vast region of the model parameter space from the LHC searches, strikingly, there is a region of this parameter space that is beyond the reach of the existing LHC searches, and doubly/singly-charged and neutral Higgses as light as 200 GeV or even lighter are still allowed by the LHC data. We study several search strategies targeting different parts of this LHC elusive parameter space at two configurations of $e^-e^+$ colliders---500 GeV and 1 TeV centre of mass energies. We find that a vast region of this parameter space could be probed with $5\sigma$ discovery with the early $e^-e^+$ colliders' data.
\end{abstract}


\maketitle

\section{\label{sec:intro}\!\!\!Introduction} \vspace{-3mm}
Among several observational and theoretical lacunae of the SM, the discovery of neutrino oscillations---necessitating the neutrinos to be massive---has provided arguably the most irrefutable reason for going beyond the SM. The widely-studied type-II see-saw model \cite{Konetschny:1977bn,Cheng:1980qt,Lazarides:1980nt,Schechter:1980gr,Mohapatra:1980yp,Magg:1980ut}, one of the three UV completions of the so-called Weinberg operator \cite{Weinberg:1979sa} at the tree-level \cite{Ma:1998dn}, extending the SM with $SU(2)_L$ triplet scalar field with hypercharge $Y\!=\!1$ offers a well-founded rationale for the observed neutrino masses and mixings. 

Copious production of the triplet-like physical Higgses, {\it viz.} doubly- and singly-charged Higgses ($H^{\pm \pm}$ and $H^\pm$) and CP-even and CP-odd neutral Higgses ($H^0$ and $A^0$), and their eventual decays to the SM fermions and bosons offer interesting ways to probe them directly at colliders. Phenomenology of these states, in particular, the doubly-charged ones, has been studied extensively at the Large Hadron Collider (LHC) \cite{Huitu:1996su,Gunion:1996pq,Chakrabarti:1998qy,Muhlleitner:2003me,Akeroyd:2005gt,Han:2007bk,delAguila:2008cj,Perez:2008ha,Akeroyd:2009hb,Akeroyd:2010ip,Melfo:2011nx,Aoki:2011pz,Akeroyd:2011zza,Chiang:2012dk,Akeroyd:2012nd,Chun:2012zu,delAguila:2013mia,Chun:2013vma,Kanemura:2013vxa,Kanemura:2014goa,Kanemura:2014ipa,kang:2014jia,Han:2015hba,Han:2015sca,Mitra:2016wpr,Ghosh:2017pxl,Antusch:2018svb,BhupalDev:2018tox,deMelo:2019asm,Primulando:2019evb,Chun:2019hce,Padhan:2019jlc,Ashanujjaman:2021txz}, $e^-e^+$ colliders \cite{Nomura:2017abh,Blunier:2016peh,Crivellin:2018ahj,Agrawal:2018pci,Rahili:2019ixf} and $e^-p$ collider \cite{Dev:2019hev,Yang:2021skb}; see Refs.~\cite{Deppisch:2015qwa,Cai:2017mow} for comprehensive reviews. The observations being consistent with the SM expectations, several LHC searches performed by the CMS and ATLAS collaborations have put stringent limits on them \cite{ATLAS:2012hi,Chatrchyan:2012ya,ATLAS:2014kca,Khachatryan:2014sta,CMS:2016cpz,CMS:2017pet,Aaboud:2017qph,CMS:2017fhs,Aaboud:2018qcu,Aad:2021lzu}. For $H^{\pm \pm}$ decaying exclusively into same-sign lepton pair, the CMS multilepton search in Ref.~\cite{CMS:2017pet} has set a limit of 535--820 GeV. The ATLAS multilepton search in Ref.~\cite{Aaboud:2017qph} has set a limit of 770--870 GeV and 450 GeV for $H^{\pm \pm}$ decaying, respectively, 100\% and 10\% into same-sign light lepton pair. For $H^{\pm \pm}$ decaying exclusively into same-sign $W$-boson pair, the ATLAS search in Ref.~\cite{Aad:2021lzu} has set a limit of 350 GeV and 230 GeV, respectively, for the pair and associated production modes. 

Understandably, the LHC searches being designed to probe specific parts of the parameter space defined by the doubly-charged Higgs mass ($m_{H^{\pm \pm}}$), the mass-splitting between the doubly- and singly-charged Higgses ($\Delta m \!=\! m_{H^{\pm \pm}} \!-\! m_{H^\pm}$) and the triplet vacuum expectation value ($v_t$), the resulted limits are not applicable for the entire parameter space. Ref.~\cite{Ashanujjaman:2021txz}, incorporating all the relevant productions and decays for the triplet-like Higgses, has derived the most stringent limit at $95\%$ confidence level on $m_{H^{\pm \pm}}$ for a vast range of $v_t$-$\Delta m$ parameter space by recasting several searches by CMS and ATLAS. It has been shown in Ref.~\cite{Ashanujjaman:2021txz} that while the triplet-like Higgses up to a few hundred GeV masses are already excluded for $\Delta m \!=\! 0$ and $\Delta m \!<\! 0$ from the LHC searches, strikingly, there is a region of the parameter space---with large enough positive $\Delta m$ and moderate $v_t$---that is beyond the reach of the existing LHC searches, and Higgses as light as 200 GeV or even lighter are still allowed by the LHC data. The challenges in probing this part of the parameter space at the LHC arise because the charged Higgses decay exclusively to the neutral ones and off-shell $W$-bosons resulting in soft hadrons or leptons, which are challenging to reconstruct at the LHC. Thus, charged Higgses' productions only enhance the production of neutral Higgses, which then decay into neutrinos or $b\bar{b},t\bar{t},ZZ,Zh,hh$, thereby resulting in final states that are challenging to probe at the LHC owing to the towering SM backgrounds. However, future lepton colliders \cite{CLICPhysicsWorkingGroup:2004qvu,Baer:2013cma,FCC:2018evy,CEPCStudyGroup:2018ghi} are expected to have better prospects for probing this region of parameter space owing to a cleaner environment. This work studies several search strategies targeting different parts of the above-mentioned LHC elusive parameter space at future $e^-e^+$ colliders. To this end, we consider two configurations of $e^-e^+$ colliders---500 GeV and 1 TeV centre of mass energies ($\sqrt{s}$).

The rest of this work is structured as follows. In Section~\ref{sec:model}, we briefly discuss the type-II see-saw model and the productions and decays of the triplet-like Higgses. We perform a comprehensive collider analysis for them in several final states at the 500 GeV and 1 TeV $e^-e^+$ colliders in Section~\ref{sec:collider}. Finally, we summarise in Section~\ref{sec:conclusion}.

\vspace{-4mm} \section{\label{sec:model}\!\!\!The Higgs triplet} \vspace{-3mm}
In addition to the SM field content, the type-II see-saw model employs a $SU(2)_L$ triplet scalar field with $Y\!=\!1$:
\[
\Delta = \begin{pmatrix} \Delta^+/\sqrt{2} & \Delta^{++} \\ \Delta^0 & -\Delta^+/\sqrt{2} \end{pmatrix}.
\]
The scalar potential involving $\Delta$ and the SM Higgs doublet $\Phi = \begin{pmatrix} \Phi^+ \!&\! \Phi^0 \end{pmatrix}^T $ is given by \cite{Arhrib:2011uy}
\begin{align*}
V(\Phi,\Delta) =& -m_\Phi^2{\Phi^\dagger \Phi} + \frac{\lambda}{4}(\Phi^\dagger \Phi)^2 + m_\Delta^2{\rm Tr}(\Delta^{\dagger}{\Delta}) 
\\
& + [\mu(\Phi^T{i}\sigma^2\Delta^\dagger \Phi)+{\rm h.c.}] + \lambda_1(\Phi^\dagger \Phi){\rm Tr}(\Delta^{\dagger}{\Delta})
\\
&  + \lambda_2[{\rm Tr}(\Delta^{\dagger}{\Delta})]^2 + \lambda_3{\rm Tr}[(\Delta^{\dagger}{\Delta})^2] + \lambda_4{\Phi^\dagger \Delta \Delta^\dagger \Phi},
\end{align*}
where $m_\Phi^2, m_\Delta^2$ and $\mu$ are the mass parameters, $\lambda$ and $\lambda_i$ ($i\!=\!1,\dots,4$) are the dimensionless quartic couplings. The neutral components of $\Phi$ and $\Delta$ can be parametrised as $\Phi^0=(v_d+h+iZ_1)\!/\sqrt{2}$ and $\Delta^0=(v_t+\xi+iZ_2)\!/\sqrt{2}$, where $v_d$ and $v_t$ are their respective vacuum expectation values (VEVs) with $\sqrt{v_d^2+2v_t^2}=246$ GeV. After the electroweak symmetry is broken, the degrees of freedom carrying identical electric charges mix, thereby resulting in several physical Higgs states: 
\begin{enumerate}
\item the neutral states $\Phi^0$ and $\Delta^0$ mix into two CP-even states ($h$ and $H^0$) and two CP-odd states ($G^0$ and $A^0$), 
\item the singly-charged states $\Phi^\pm$ and $\Delta^\pm$ mix into two mass states $G^\pm$ and $A^\pm$, 
\item the doubly-charged gauge state $\Delta^{\pm \pm}$ is aligned with its mass state $H^{\pm \pm}$.
\end{enumerate}
The mass states $G^0$ and $G^\pm$ are the so-called {\it would-be} Nambu-Goldstone bosons eaten by the longitudinal modes of $Z$ and $W^\pm$, and the rest of them are massive with $h$ being identified as the 125-GeV resonance observed at the LHC.

The Yukawa interaction $Y^{\nu}_{ij} L^T_i C i \sigma^2 \Delta L_j$ of the triplet Higgs with the SM lepton doublet leads to the non-zero masses for the neutrinos after the electroweak symmetry breaking ($Y^\nu$ is a $3\times 3$ symmetric complex matrix, $i$ and $j$ are the generation indices, and $C$ is the charge-conjugation matrix):
\[
m_\nu=\sqrt{2}Y^\nu v_t.
\]
Consequently, $Y^\nu$ is determined by the neutrino oscillation parameters up to the triplet VEV. In this work, we take the best fit values for the neutrino oscillation parameters from Ref.~\cite{Esteban:2020cvm} except for the Dirac and Majorana phases which we set to zero for simplicity.

In this model, there are only three phenomenologically relevant parameters, namely the doubly-charged Higgs mass ($m_{H^{\pm \pm}}$), the mass-splitting between the doubly- and singly-charged Higgses ($\Delta m \!=\! m_{H^{\pm \pm}} \!-\! m_{H^\pm}$) and $v_t$. The triplet-like singly-charged and neutral Higgs masses are given by
\[
m_{H^\pm} = m_{H^{\pm \pm}} -\Delta m \,\mbox{and}\, m_{H^0/A^0} \approx \sqrt{m_{H^{\pm \pm}} (m_{H^{\pm \pm}} -4\Delta m)}.
\]
For the sake of completeness, we briefly mention the relevant constraints on them.
\begin{enumerate}
\item The value of the $\rho$ parameter from the electroweak precision data (EWPD) \cite{Zyla:2020zbs} leads to an upper bound of $\mathcal{O}(1)$ GeV on $v_t$.
\item The EWPD observables, {\it viz.} $S, T$ and $U$ parameters robustly constrain the mass-splittings requiring $|\Delta m| \lesssim 40$ GeV \cite{Aoki:2012jj,Chun:2012jw,Primulando:2019evb,Das:2016bir}.
\item The upper limits on the lepton flavour violating decays $\mu^- \to e^- \gamma$ \cite{TheMEG:2016wtm} and $\mu^- \to e^+e^-e^-$ \cite{Bellgardt:1987du} tightly constrain the $v_t$--$m_{H^{\pm \pm}}$ parameter space \cite{Kakizaki:2003jk,Akeroyd:2009nu,Dinh:2012bp}: \vspace{-4mm}
\[
\qquad \,\,\, v_t \gtrsim \mathcal{O}(10^{-9}) {\rm ~GeV} \times \frac{1~ \rm TeV}{m_{H^{\pm \pm}}}.
\]
\item \vspace{-4mm} For $\Delta m=0$ and large (small) $v_t$, doubly charged scalars with masses below 420(955) GeV are excluded from the LHC searches \cite{Ashanujjaman:2021txz}. For large enough negative $\Delta m$ and moderate $v_t$, the exclusion limit extends up to 1115 GeV. However, for large enough positive $\Delta m$ and moderate $v_t$, triplet-like Higgses as light as 200 GeV or even lighter are still allowed by the LHC data.
\end{enumerate}

\begin{figure}[htb!]
\centering
\includegraphics[width=0.79\columnwidth]{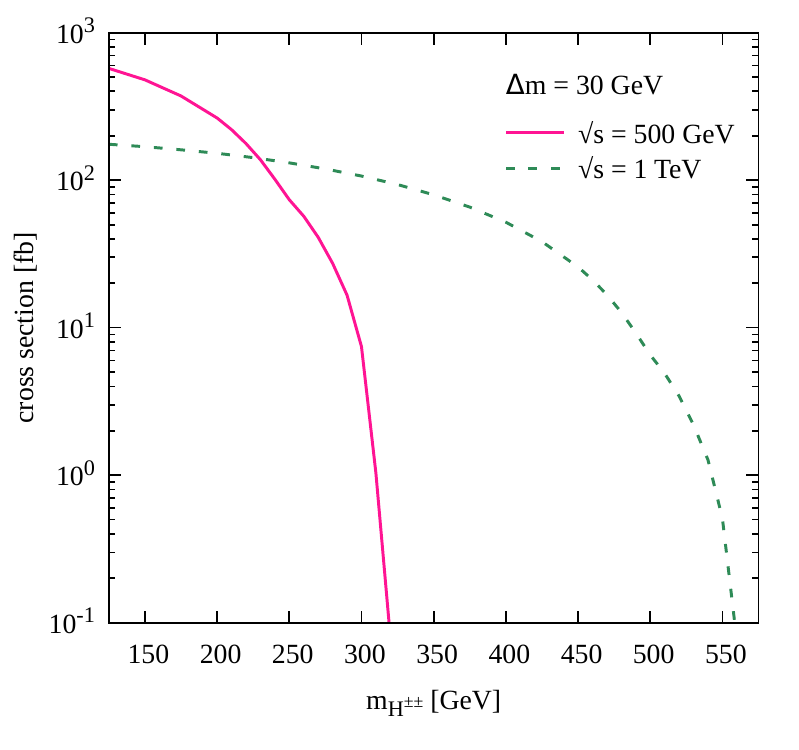}
\caption{Total production cross-sections for the triplet-like Higgses at the 500 GeV and 1 TeV $e^-e^+$ colliders.}
\label{fig:xsec}
\end{figure}

The triplet Higgses can be pair produced aplenty at $e^-e^+$ colliders through $s$-channel $\gamma/Z$ exchanges \cite{Agrawal:2018pci}:\footnote{They can also be single or pair produced via vector boson-fusion processes with two associated forward leptons at $e^-e^+$ colliders. However, we do not consider these due to their small contribution.}
\[
e^- e^+ \to H^{++} H^{--}, H^+ H^-, H^0 A^0.
\]
We evaluate the leading order production cross-sections using the \texttt{SARAH} \cite{Staub:2013tta,Staub:2015kfa} generated \texttt{UFO} \cite{Degrande:2011ua} modules in \texttt{MadGraph} \cite{Alwall:2011uj,Alwall:2014hca}. Fig.~\ref{fig:xsec} shows the total production cross-sections for the triplet-like Higgses as a function of $m_{H^{\pm \pm}}$ for $\Delta m=30$ GeV at both the 500 GeV and 1 TeV $e^-e^+$ colliders.

\begin{figure}[]
\centering
\includegraphics[width=0.95\columnwidth]{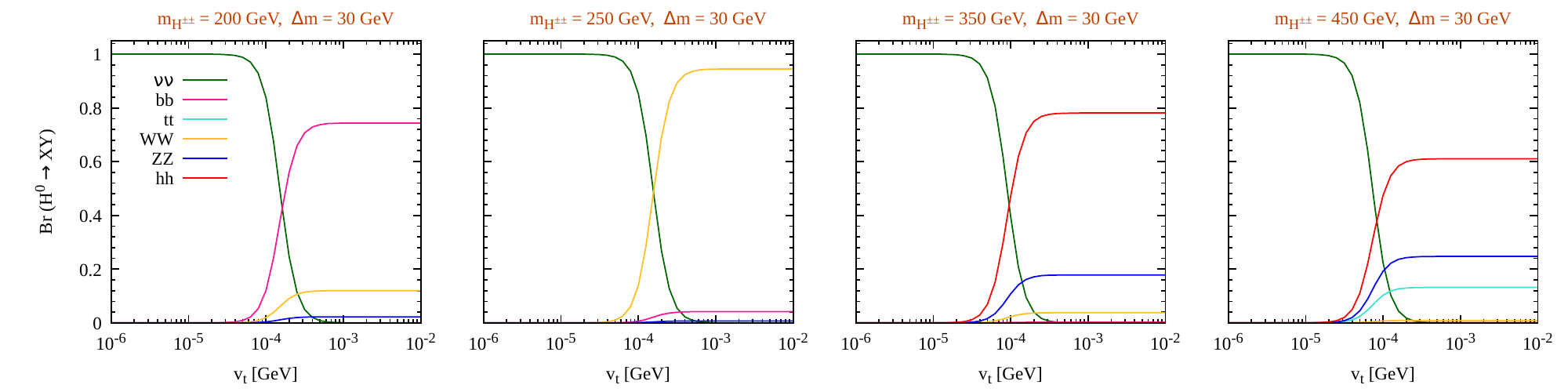}
\includegraphics[width=0.95\columnwidth]{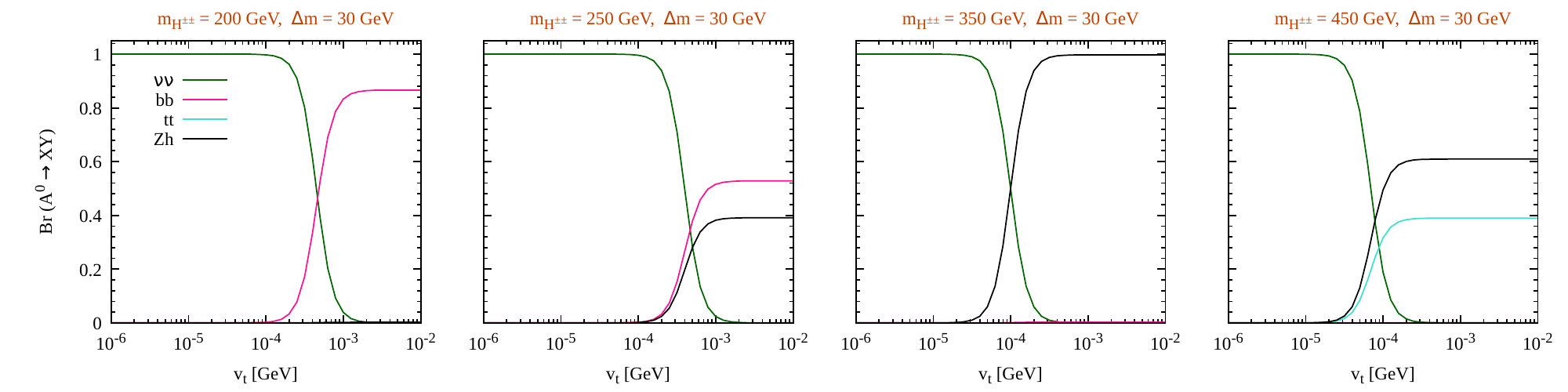}
\caption{Branching fractions for the triplet-like CP-even (top panel) and CP-odd (bottom pannel) neutral Higgses.}
\label{fig:br}
\end{figure}

After being produced, the triplet-like Higgses decay either to a lighter triplet-like Higgs and an off-shell $W$-boson or a pair of SM particles. For the model parameter space of our interest---with large enough positive $\Delta m$ and moderate $v_t$, $H^{\pm \pm}$ and $H^\pm$ undergo the former decay, thereby enhancing the productions of $H^0$ and $A^0$ effectively. Finally, $H^0$ and $A^0$ decay into a pair of neutrinos or/and $b\bar{b},t\bar{t},ZZ,Zh,hh$. In Fig.~\ref{fig:br}, we present their branching fractions as a function of $v_t$ for different values of $m_{H^{\pm \pm}}$---200, 250, 350 and 450 GeV with $\Delta m=30$ GeV. As these plots suggest, the dominance of a decay mode over the others depends, naturally, on their mass and $v_t$. For instance, for $v_t > \mathcal{O}(10^{-4})$ GeV, while a light $H^0/A^0$ decay dominantly into $b\bar{b}$, the heavier $A^0(H^0)$ decay into $Zh$ ($WW$ and $hh$). For $v_t < \mathcal{O}(10^{-4})$ GeV, they decay exclusively into a pair of neutrinos irrespective of their mass.

\vspace{-4mm} \section{\label{sec:collider}\!\!\!Collider phenomenology} \vspace{-3mm}
Abundant production of the triplet-like Higgses followed by their eventual decays to the SM particles will lead to various final state signatures at $e^-e^+$ colliders. Led the way by the decay patterns of $H^0$ and $A^0$, we define several signal regions (SRs) targeting different parts of the above-mentioned parameter space, see Table~\ref{SRs} (and the corresponding benchmark points are summarised in Table~\ref{BPs}). Before going into the SR-specific selection, we briefly describe the reconstruction and selection of various physics objects.
\newpage

\onecolumngrid

\twocolumngrid

\begin{table}[]
\scalebox{0.83}{
\begin{tabular}{p{6mm}l *{3}{l} l l}
\toprule
\multirow{2}{*}[-2pt]{SR} & \multicolumn{3}{c}{parameter space of interest} & \multirow{2}{*}[-2pt]{Final state of interest} & \multirow{2}{*}[-2pt]{$\sqrt{s}$} \\
\cmidrule(lr){2-4}
& \vspace{-0.2cm} $v_t$ & $\Delta m$ & $m_{H^{\pm \pm}}$ & & & \\
& \vspace{-0.25cm} (GeV) & (GeV) & (GeV) & & (GeV) \\
\midrule
SR1 & $\gtrsim \mathcal{O}(10^{-4})$ & 30 & $\lesssim 250$ & $\geq 2 b$-jets + anything & 500 \\
SR2 & $\mathcal{O}(10^{-4})$ -- $\mathcal{O}(10^{-3})$ & 30 & $\lesssim 290$ & $\geq 3$ jets + $p_T^{\rm miss}$ & 500 \\
SR3 & $\lesssim \mathcal{O}(10^{-4})$ & 30 & $\lesssim 250(500)$ & soft leptons/jets & 500(1000) \\
SR4 & $\gtrsim \mathcal{O}(10^{-4})$ & 30 & 250--550 & $\geq 7$ jets & 1000 \\
\bottomrule
\end{tabular}
}
\caption{\label{SRs} SRs targeting different parts of the parameter space.}
\end{table}

\begin{table}[h]
\begin{tabular}{p{7mm}l *{4}{l}}
\toprule
\multirow{2}{*}[-2pt]{SR} & \multicolumn{4}{c}{Benchmark points} \\
\cmidrule(lr){2-5}
& name & $m_{H^{\pm \pm}}$ & $\Delta m$ & $v_t$ \\
& & (GeV) & (GeV) & (GeV) \\
\midrule
SR1 & {\it BP1} & 230 & 30 & $3\times 10^{-4}$ \\
SR2 & {\it BP2} & 260 & 30 & $3\times 10^{-4}$ \\
SR3 & {\it BP3} ({\it BP4}) & 240 (425) & 30 & $10^{-5}$ \\
SR4 & {\it BP5} & 375 & 30 & $3\times 10^{-4}$ \\
\bottomrule
\end{tabular}
\caption{\label{BPs} Benchmark points used for different SRs.}
\end{table}

\vspace{-4mm} \subsection{\!\!\!Object reconstruction and selection} \vspace{-3mm}
We use \texttt{MadGraph} \cite{Alwall:2011uj,Alwall:2014hca} to simulate parton-level events for both signals and backgrounds. We pass those events into \texttt{PYTHIA} \cite{Sjostrand:2014zea} to simulate subsequent decays for the unstable particles, initial and final state radiations, showering, fragmentation and hadronisation. Finally, we pass them into \texttt{Delphes} \cite{deFavereau:2013fsa} for simulating detector effects as well as reconstructing various physics objects, {\it viz.} photons, electrons, muons and jets. Jets are reconstructed using the {\it anti-k$_T$ algorithm} \cite{Cacciari:2008gp} with a characteristic radius 0.4 in \texttt{FastJet} \cite{Cacciari:2011ma}. Jets (leptons, {\it i.e.} electrons and muons, and photons) are required to be within the pseudorapidity range $|\eta|<2.4(2.5)$ and have a transverse momentum $p_T > 10(5)$ GeV. Further, muons (photons and electrons) are required to be isolated, and this is ensured by demanding the scalar sum of the $p_T$s of all other objects lying within a cone of radius 0.5 around it to be smaller than 15\%(12\%) of its $p_T$. Such stringent isolation requirements significantly suppress the reducible backgrounds. Finally, the missing transverse momentum vector $\vec p_T^{\rm\,\,miss}$ (with magnitude $p_T^{\rm miss}$) is estimated from the momentum imbalance in the transverse direction associated to all reconstructed objects in an event.

\vspace{-4mm} \subsection{\!\!\!SM Backgrounds} \vspace{-3mm}
While different SM processes serve as the main background for different SRs, for the sake of completeness, we consider all the relevant backgrounds across the SRs. These include diboson ($VV$ with $V$ denoting the gauge bosons), triboson ($VVV$) and tetraboson ($VVVV$) productions, Higgsstrahlung processes ($Vh$, $VVh$, $Vhh$, $t\bar{t}h$), multi-top production ($t\bar{t}$, $t\bar{t}t\bar{t}$), top-pair production in association with gauge bosons ($t\bar{t}V$, $t\bar{t}VV$), dilepton production and multi-jet production.

\vspace{-4mm} \subsection{\!\!\!SR-specific event selection} \vspace{-3mm}
We now briefly discuss the SR-specific selection criteria that would significantly suppress the background without impinging much on the signal. To achieve this, we use various kinematic distributions as a guiding premise.

\vspace{-4mm} \subsubsection{\!\!\!SR1: \small{$v_t \gtrsim \mathcal{O}(10^{-4})$, $\Delta m \sim 30$, $m_{H^{\pm\pm}} \lesssim 250~{\rm GeV}$}} 
\begin{figure}[htb!]
\centering
\includegraphics[width=0.49\columnwidth]{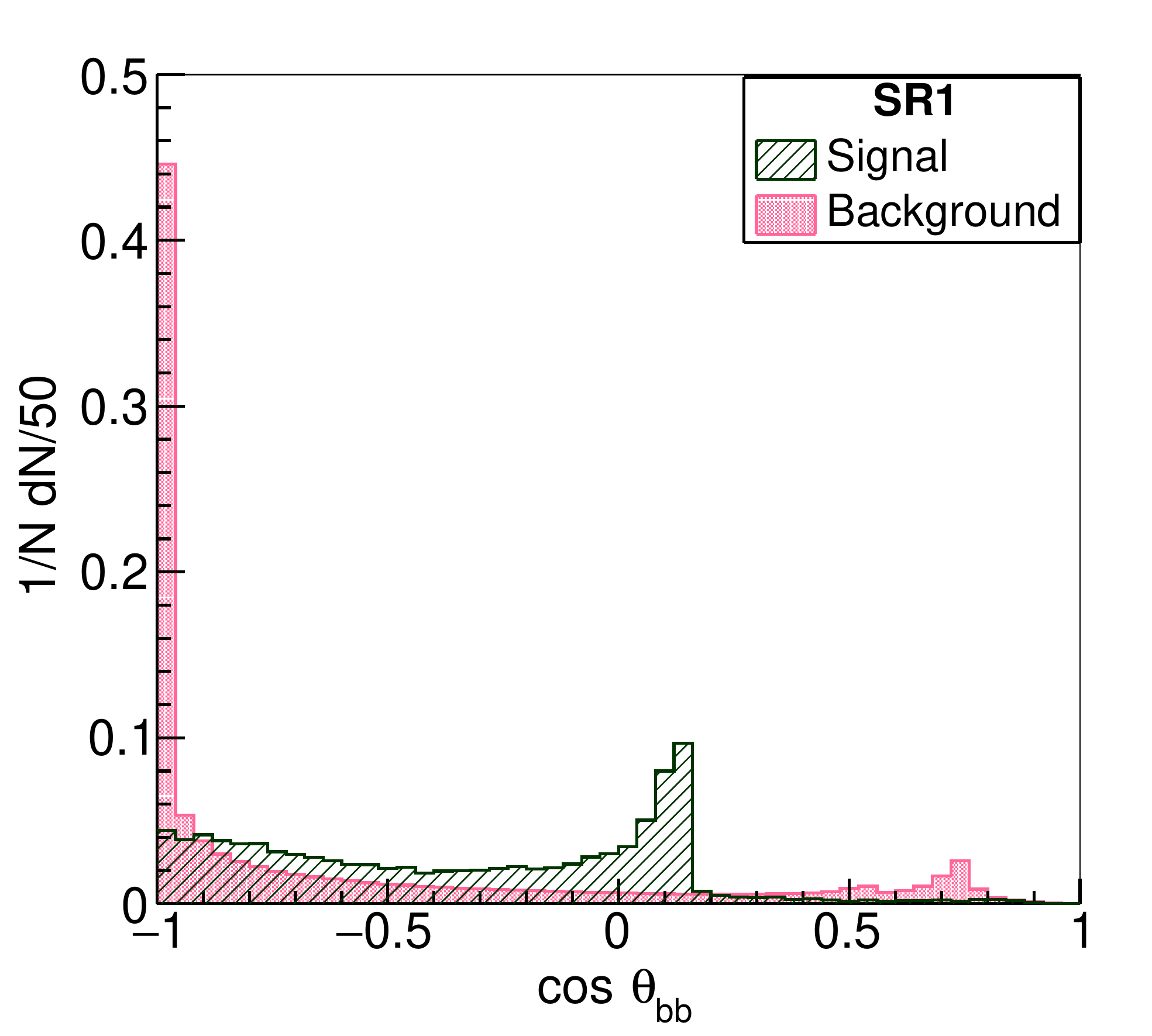}
\includegraphics[width=0.49\columnwidth]{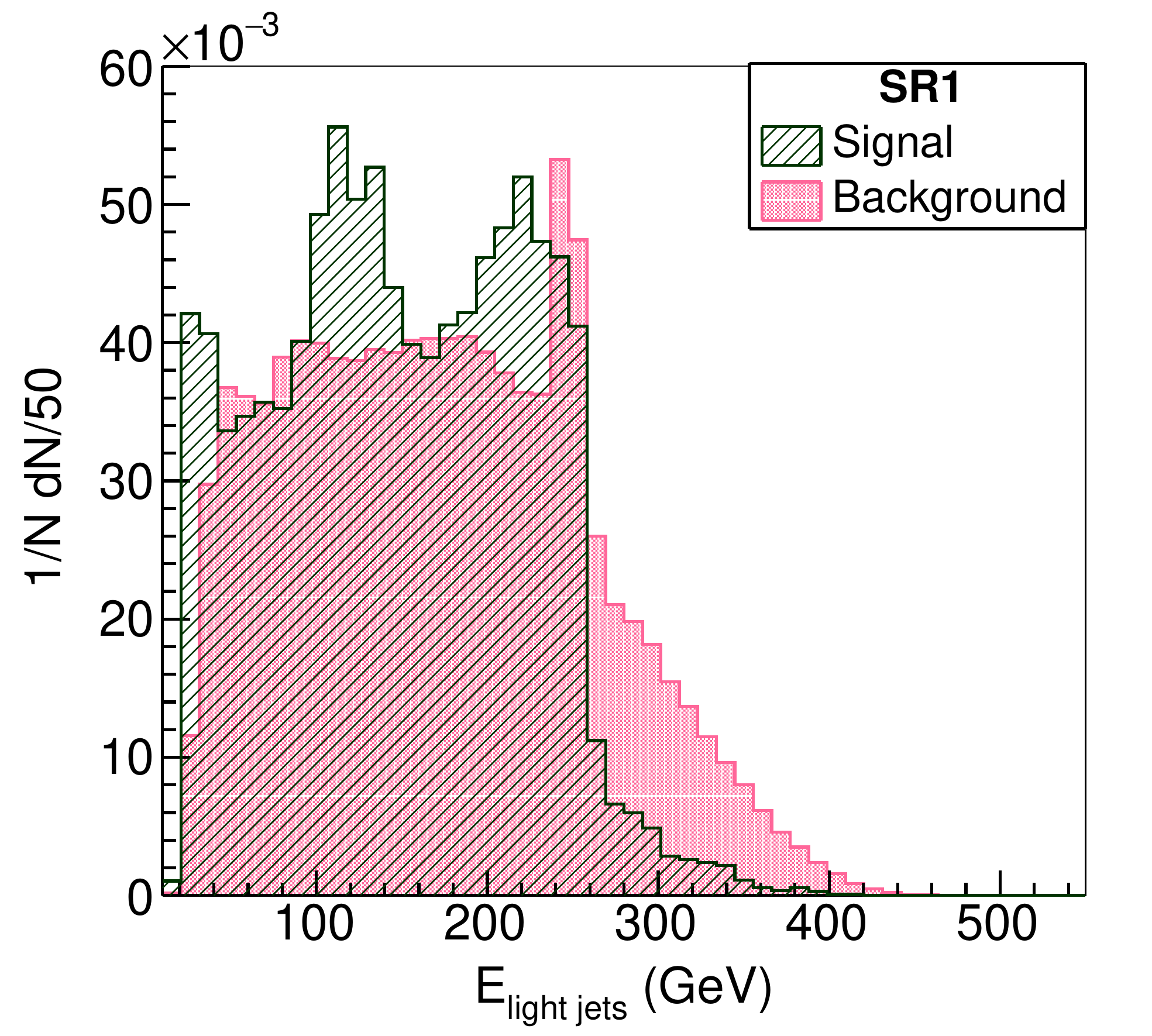}
\caption{Normalised kinematic distributions for the signal ({\it BP1}) and background at the 500 GeV $e^-e^+$ collider. Left: $\cos\theta_{bb}$ and right: $E_{\rm light~\!jets}$.}
\label{fig:SR1}
\end{figure}

\vspace{-3mm}
In this SR, $A^0$ dominantly decays into $b\bar{b}$, and $H^0$ decays to $\nu\nu$, $b\bar{b}$ or $WW$ so that the final state includes at least two $b$-jets in addition to other jets or leptons if any. We require at least two of the jets to be $b$-tagged. The invariant mass distribution of these two $b$-tagged jets are expected to peak at $m_{H^0/A^0}$. The SM processes $t\bar{t}$ and $b\bar{b}$ serve as the main irreducible backgrounds for this final state. In Fig.~\ref{fig:SR1}, we display two normalised kinematic distributions for the signal and background events at the 500 GeV $e^-e^+$ collider. The signal events are shown for a benchmark point {\it BP1}: $m_{H^{\pm \pm}}=230$ GeV, $\Delta m=30$ GeV and $v_t \sim 3\times 10^{-4}$ GeV. The left plot shows the distributioin for the angle between the two $b$-tagged jets ($\cos\theta_{bb}$). In case of more than two $b$-tagged jets, the pair with maximum separation in the azimuth plane is considered. The background boasts a peak around $\cos\theta_{bb} = -1$ with the most dominant contribution coming from $b\bar{b}$ events with the pair of $b$-jets emanating back-to-back. Displayed in the right plot is the distribution of the sum of all non $b$-tagged jet energies ($E_{\rm light~\!jets}$). To improve the signal-to-background ratio, we impose the following selection cuts:
\[
\cos\theta_{bb} \in [-0.96,0.4] {\rm ~and~} E_{\rm light~\!jets} < 250 {\rm ~GeV}.
\]

\begin{figure}[htb!]
\centering
\includegraphics[width=0.6\columnwidth]{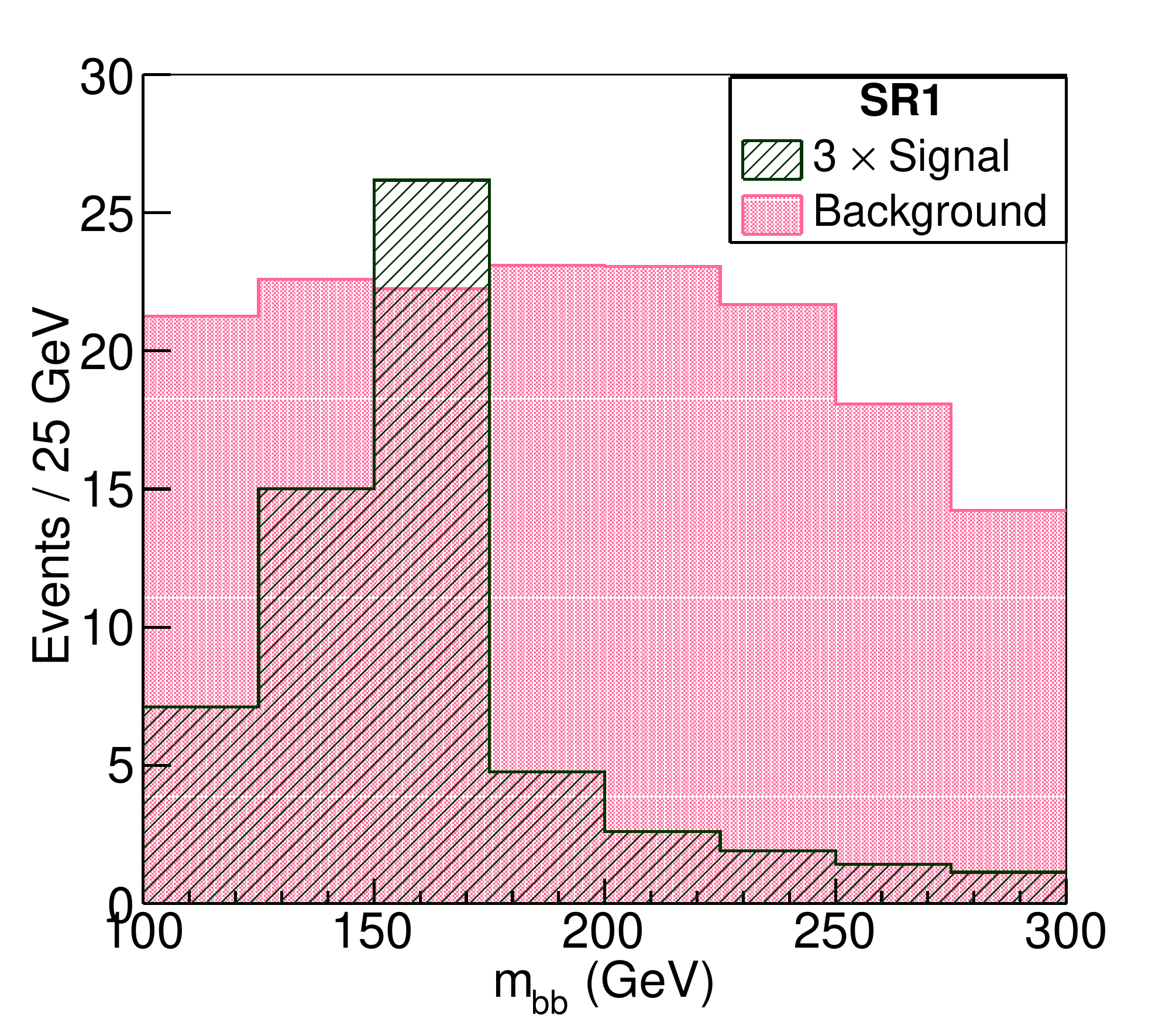}
\caption{$m_{bb}$ distribution for the signal ({\it BP1}) and background. The events are weighted for 1 fb$^{-1}$ luminosity at the 500 GeV $e^-e^+$ collider.}
\label{fig:SR11}
\end{figure}

The sensitivity of this search is increased by dividing the selected events into 8 bins in the range [100,300] GeV using the invariant mass of the two $b$-tagged jets ($m_{bb}$) as the final discriminating variable between the signal and background (see Fig.~\ref{fig:SR11}). As we expected, the $m_{bb}$ distribution peaks in the 150--175 GeV bin, thereby reconstructing the triplet-like neutral Higgses.

\vspace{-4mm} \subsubsection{\!\!\!\!SR2: \small{$v_t \sim \mathcal{O}(10^{-4})$--$\mathcal{O}(10^{-3})$, $\Delta m \sim 30$, $m_{H^{\pm\pm}} \lesssim 290~{\rm GeV}$}} 
\begin{figure}[htb!]
\centering
\includegraphics[width=0.32\columnwidth]{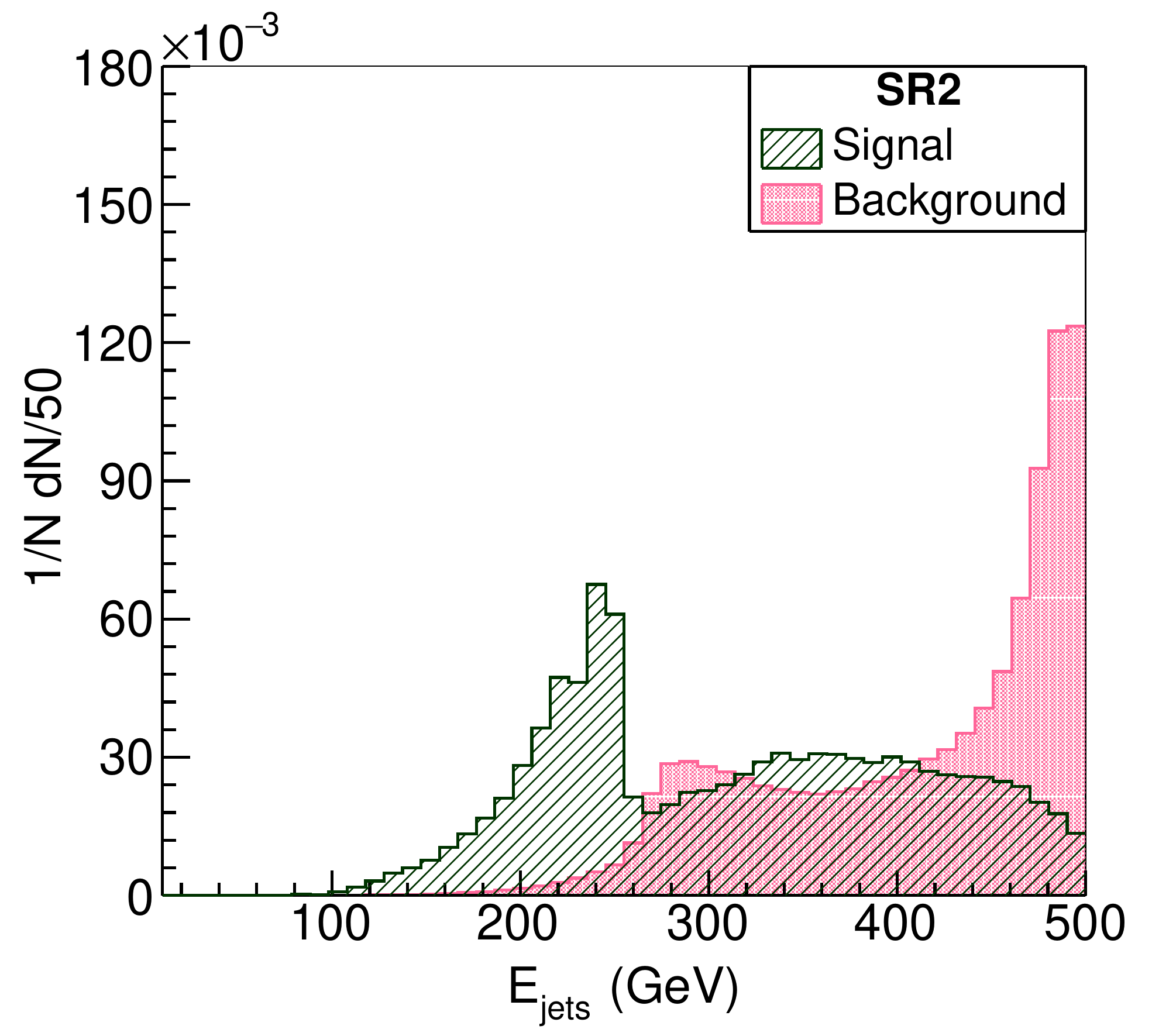}
\includegraphics[width=0.32\columnwidth]{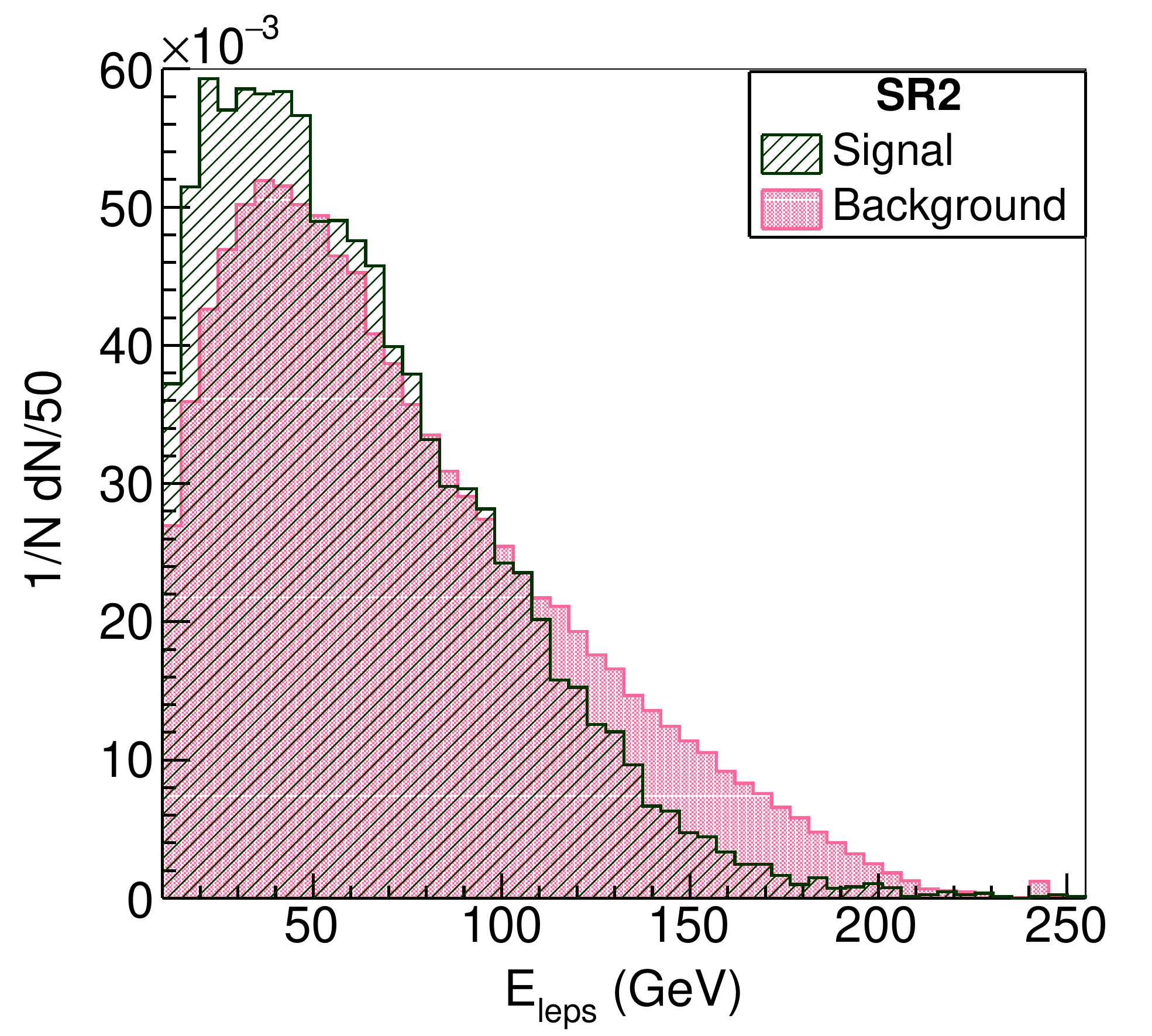}
\includegraphics[width=0.32\columnwidth]{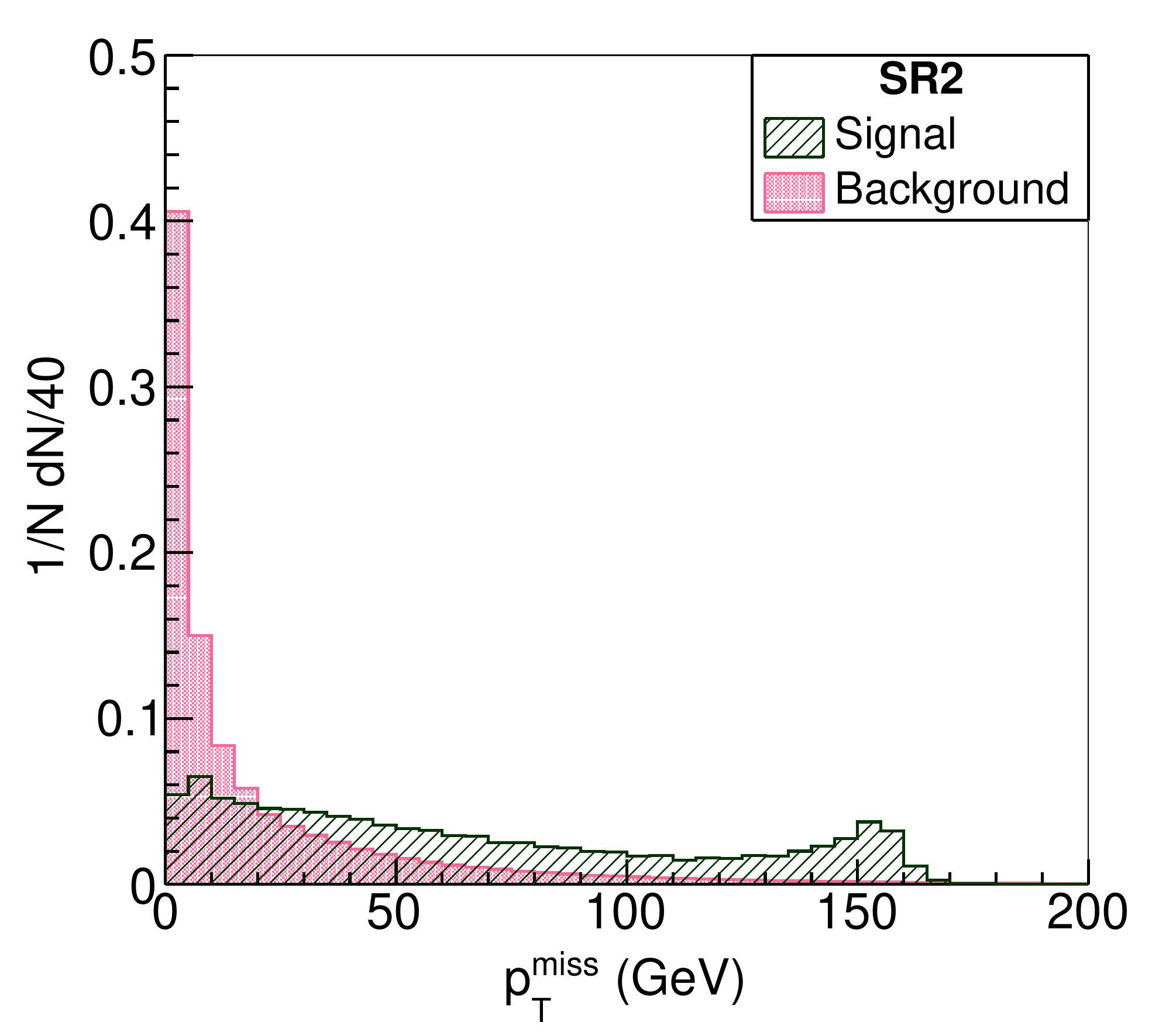}
\caption{Normalised kinematic distributions for the signal ({\it BP2}) and background at the 500 GeV $e^-e^+$ collider. Left: $E_{\rm jets}$, middle: $E_{\rm leps}$ and right: $p_T^{\rm miss}$.}
\label{fig:SR2}
\end{figure}

\vspace{-3mm}
In this SR, one of $A^0$ and $H^0$ decays into $\nu\nu$, and the other decays to $WW$ or $Zh$. We require at least three jets 
in the final state. To suppress the background contributions from $t\bar{t}$ and $b\bar{b}$ processes, we apply a $b$-jet veto. We display some normalised kinematic distributions for the signal and background events at the 500 GeV $e^-e^+$ collider in Fig.~\ref{fig:SR2}. The signal events are shown for a benchmark point {\it BP2}: $m_{H^{\pm \pm}}=260$ GeV, $\Delta m=30$ GeV and $v_t \sim 3\times 10^{-4}$ GeV. The left (middle) plot shows the distribution for the sum of all jet (lepton) energies, $E_{\rm jets}$ ($E_{\rm leps}$). As for the $E_{\rm jets}$ distribution for the background events, it is almost a monotonically rising one, peaking at $\sqrt{s}$, with most of the contributions coming from $WW$ and $t\bar{t}$ productions. Whereas for the signal events, it is a wide one, extending up to $\sqrt{s}$ with a peak at $\sqrt{s}/2$. Also displayed, in the right plot, is the $p_T^{\rm miss}$ distribution. For the signal, this distribution is almost flat, extending beyond 150 GeV. This is effectuated by one of the two triplet-like scalars' decay into hadronic final state, with the other decaying invisibly. On the contrary, for the background events, it is almost a monotonically falling one, falling sharply at very low $p_T^{\rm miss}$. Guided by these kinematic distributions, we impose the following selection cuts to improve the signal-to-background ratio:
\[
E_{\rm jets} < 260 {\rm ~GeV}, E_{\rm leps} < 120 {\rm ~GeV ~and~} p_T^{\rm miss} > 40 {\rm ~GeV}.
\]
Though the adoption of the cut on $E_{\rm jets}$ impinges on the signal strength, this significantly increases the signal-to-background ratio, owing to a much diminished background.

\begin{figure}[htb!]
\centering
\includegraphics[width=0.6\columnwidth]{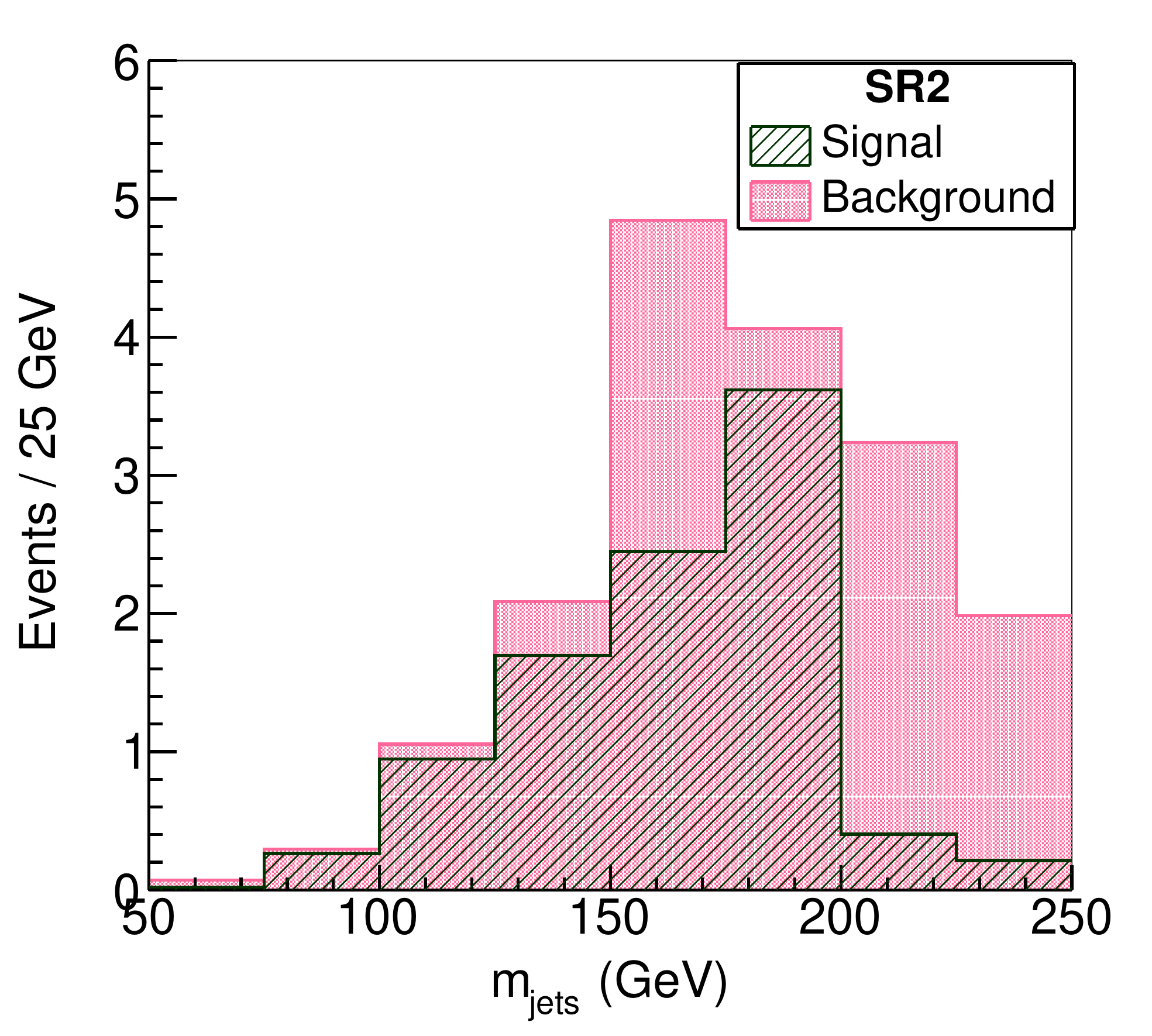}
\caption{$m_{\rm jets}$ distribution for the signal ({\it BP2}) and background. The events are weighted for 1 fb$^{-1}$ luminosity at the 500 GeV $e^-e^+$ collider.}
\label{fig:SR21}
\end{figure}

To enhance the sensitivity of this search further, the selected events are distributed over 8 bins in the range [50,250] GeV using the invariant mass of all jets, $m_{\rm jets}$ (see Fig.~\ref{fig:SR21}).

\vspace{-4mm} \subsubsection{\!\!\!SR3: \small{$v_t \lesssim \mathcal{O}(10^{-4})$, $\Delta m \sim 30$ and $m_{H^{\pm\pm}} \lesssim 500~{\rm GeV}$}} 
\begin{figure}[]
\centering
\includegraphics[width=0.48\columnwidth]{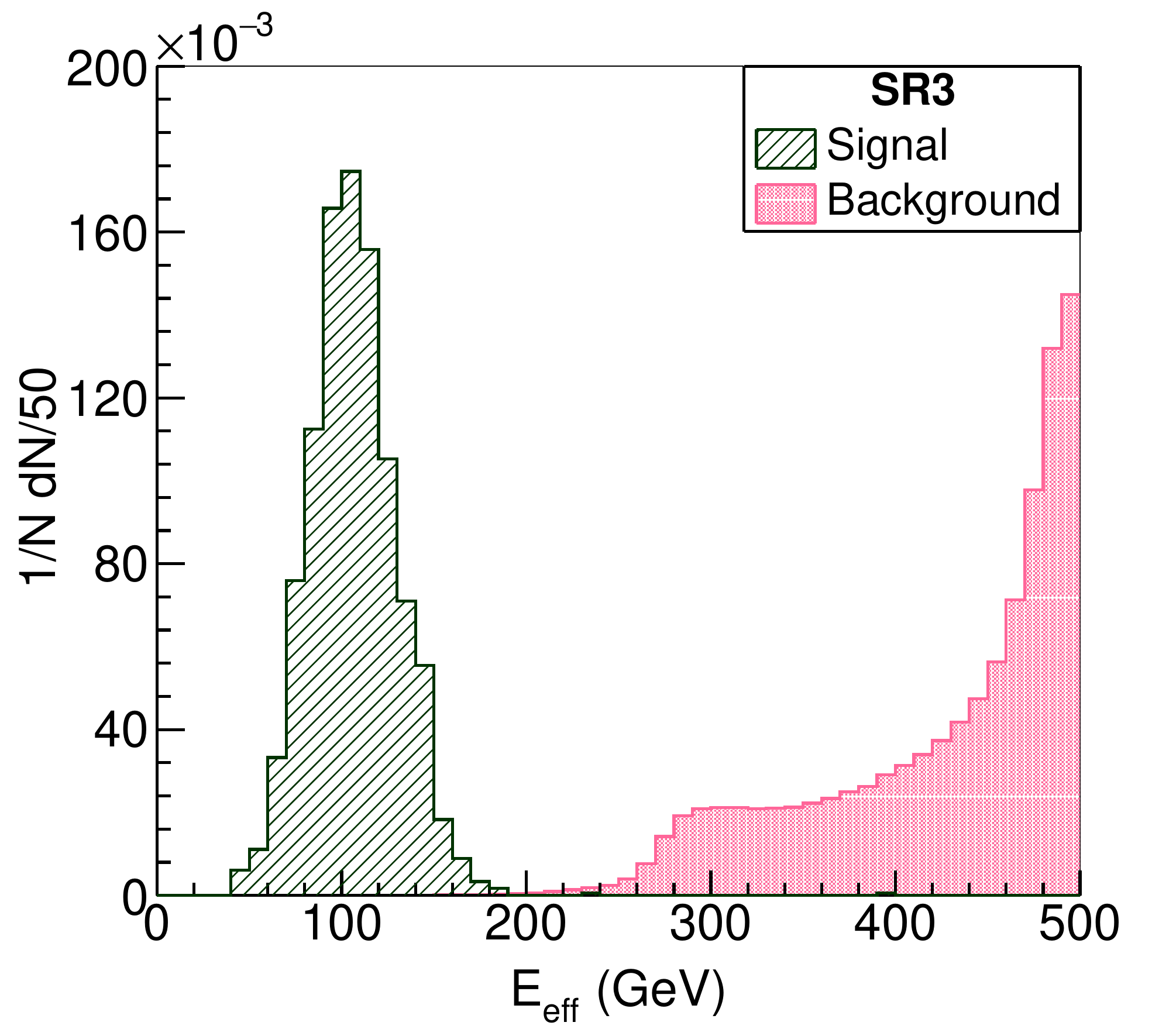}
\includegraphics[width=0.48\columnwidth]{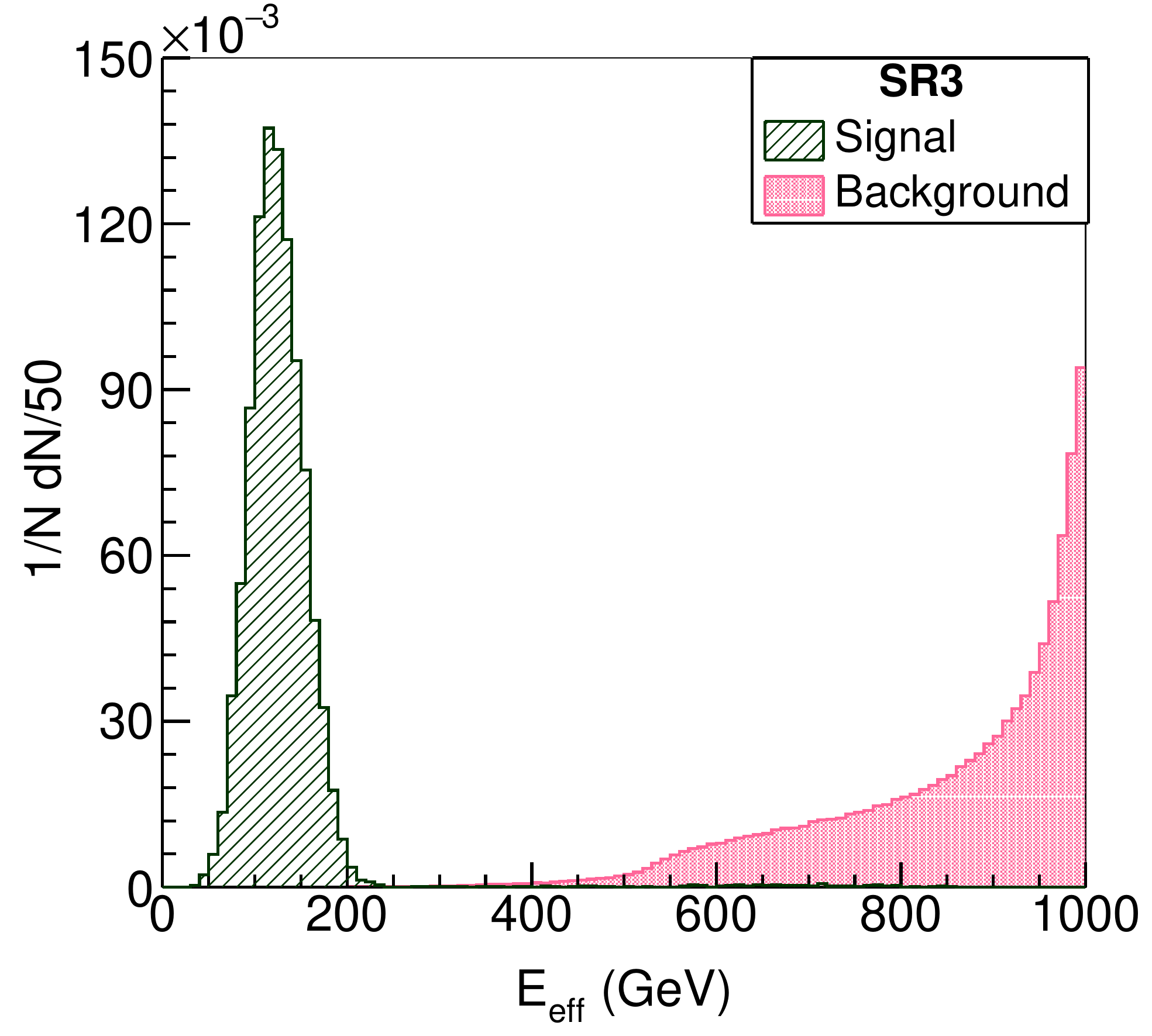}
\caption{Normalised $E_{\rm eff}$ distribution for the signal (left: {\it BP3}, right: {\it BP4}) and background at the 500 GeV (left) and 1 TeV (right) $e^-e^+$ colliders.}
\label{fig:SR3}
\end{figure}

\begin{figure}[]
\centering
\includegraphics[width=0.48\columnwidth]{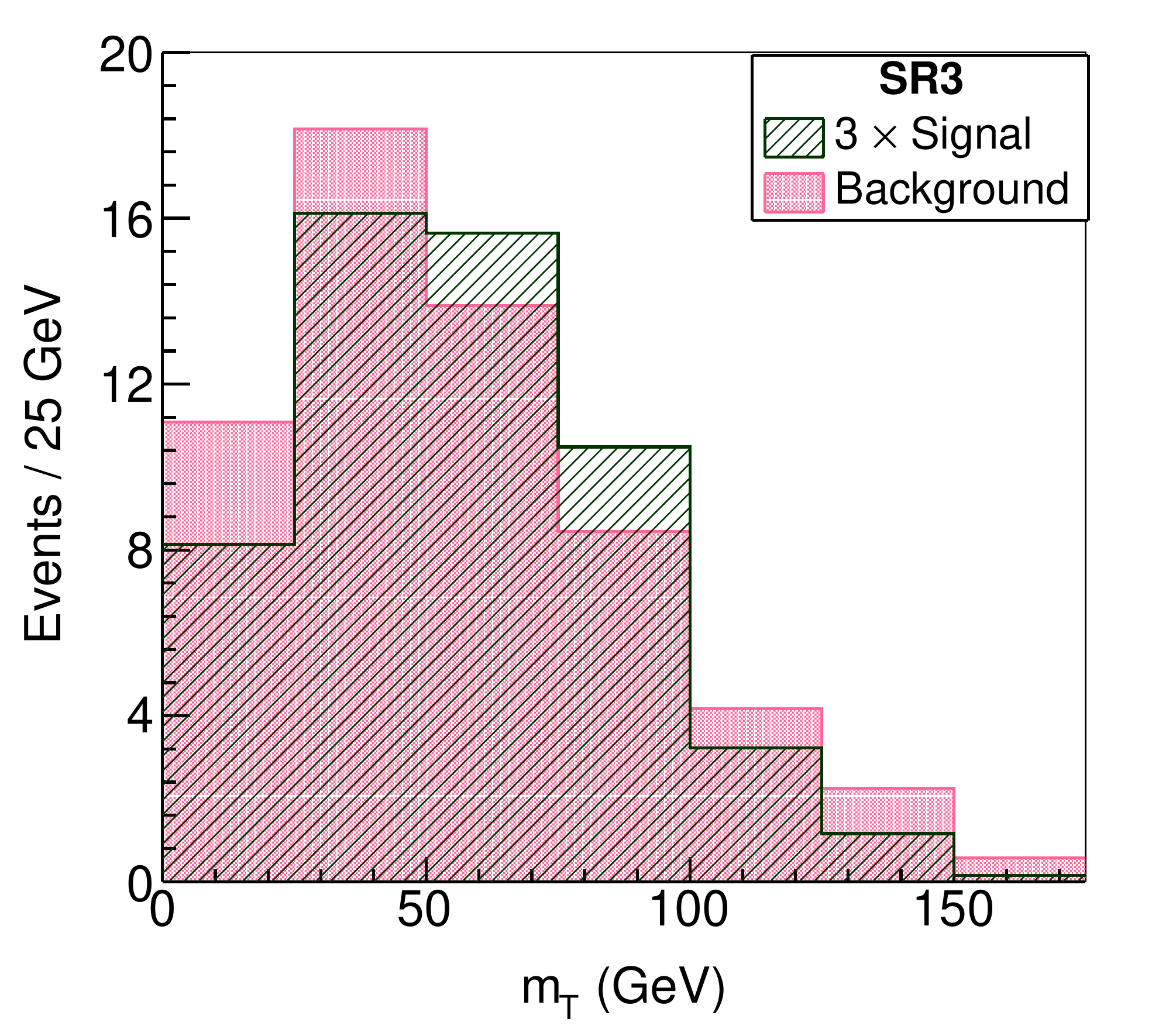}
\includegraphics[width=0.48\columnwidth]{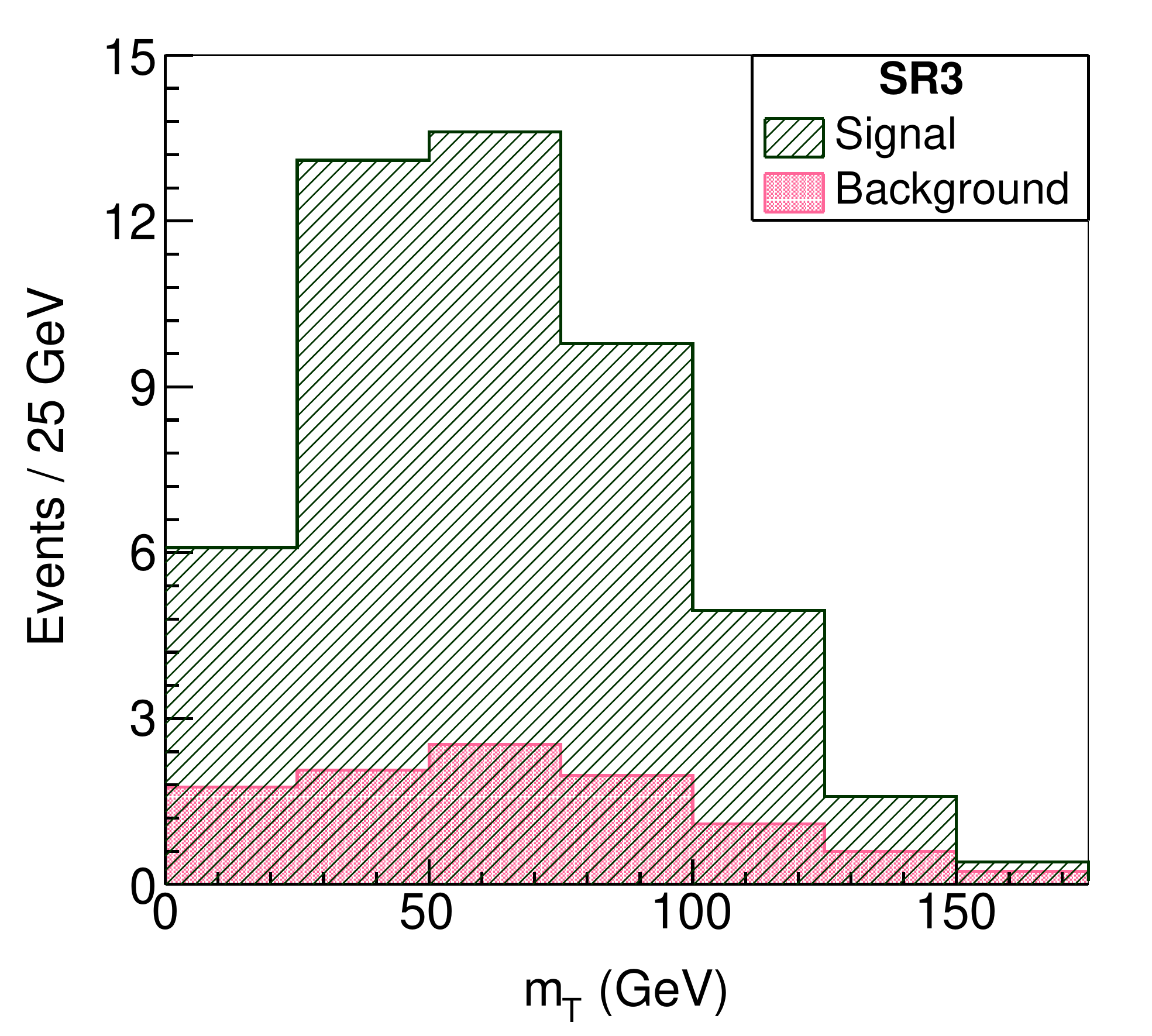}
\caption{$m_T$ distribution for the signal (left: {\it BP3}, right: {\it BP4}) and background. The events are weighted for 10 fb$^{-1}$ luminosity at the 500 GeV (left) and 1 TeV (right) $e^-e^+$ colliders.}
\label{fig:SR31}
\end{figure}

\vspace{-3mm}
In this SR, $H^{\pm \pm}$ and $H^\pm$ decay to $H^0/A^0$ and off-shell $W$-bosons. While $H^0$ and $A^0$ decay into neutrinos, the off-shell $W$-bosons decay into soft leptons/jets, thereby resulting into soft leptons/jets plus $p_T^{\rm miss}$ in the final state. We require at least three soft leptons and/or jets in the final state. In Fig.~\ref{fig:SR3}, we display the normalised kinematic distributions for $E_{\rm eff} = E_{\rm jets} + E_{\rm leps} + p_T^{\rm miss}$ for the signal and background events at the 500 GeV and 1 TeV $e^-e^+$ colliders. The signal events are shown for a benchmark point {\it BP3} ({\it BP4}): $m_{H^{\pm \pm}}=240 (425)$ GeV, $\Delta m=30$ GeV and $v_t \sim 10^{-5}$ GeV at the 500 GeV (1 TeV) collider. For the signal, this distribution boasts a peak around 100 GeV, thereafter falling sharply. This is occasioned by the softness of the leptons/jets stemming from the off-shell $W$-bosons. On the other hand, for the background events, it is almost a monotonically rising one, peaking at $\sqrt{s}$, with bulk of the contributions coming from $WW$, $ZZ$ and $t\bar{t}$ productions. Guided by these distributions, we apply the following selection cut to ameliorate the signal-to-background ratio:
\[
E_{\rm eff} < 200 (250) \mbox{~GeV for the 500 GeV (1 TeV) collider}.
\]
Finally, the selected events are distributed over 7 bins in the range [0,175] GeV using the transverse mass $m_T$ (see Fig.~\ref{fig:SR31}), where $m_T$ is defined as
\[
m_T^2 = 2 p_T^{\rm miss} p_T^{\rm vis} \left(1-\cos\Delta\phi_{\vec p_T^{\rm\,\,miss},\vec p_T^{\rm\,\,vis}}\right),
\]
where $\vec p_T^{\rm\,\,vis}$ (with magnitude $p_T^{\rm vis}$) is the vector sum of the transverse momenta of all leptons and jets, and $\Delta\phi_{\vec p_T^{\rm\,\,miss},\vec p_T^{\rm\,\,vis}}$ is the azimuthal separation between $\vec p_T^{\rm\,\,miss}$ and $\vec p_T^{\rm\,\,vis}$.

\vspace{-4mm} \subsubsection{\!\!\!SR4: \small{$v_t \gtrsim \mathcal{O}(10^{-4})$, $\Delta m \sim 30$, $m_{H^{\pm\pm}} \sim 250$--$550~{\rm GeV}$}} 
\begin{figure}[]
\centering
\includegraphics[width=0.24\columnwidth]{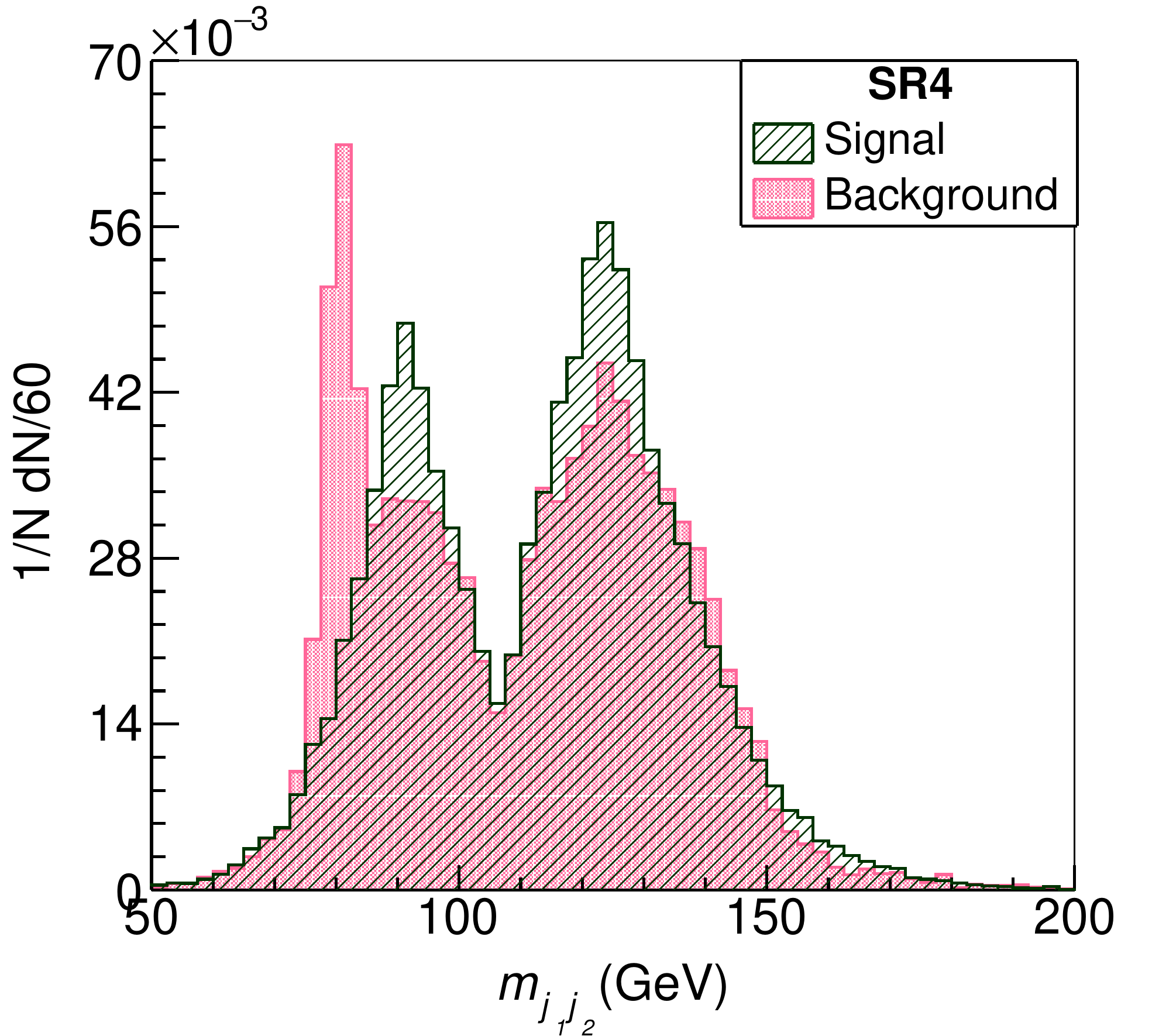}
\includegraphics[width=0.24\columnwidth]{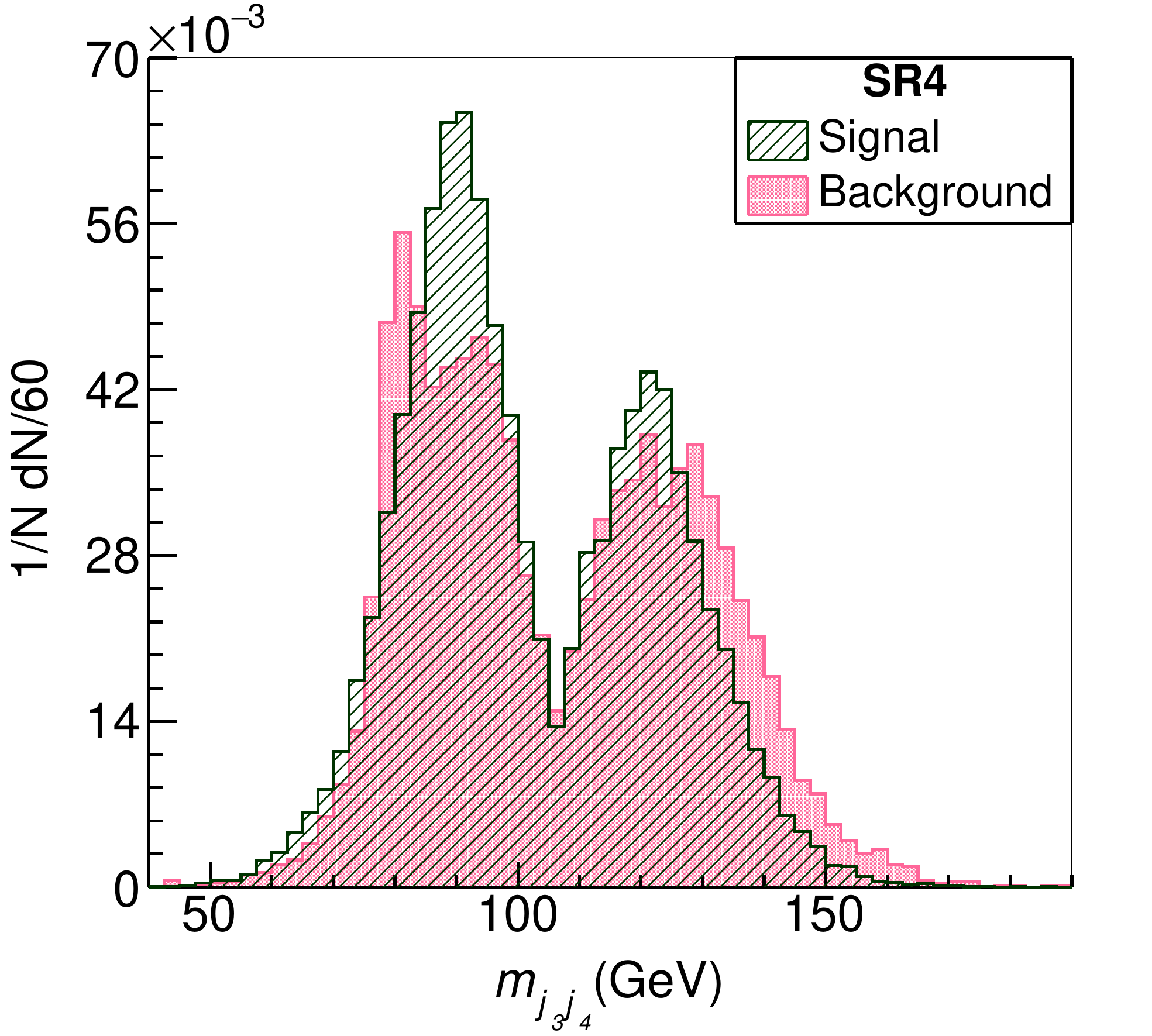}
\includegraphics[width=0.24\columnwidth]{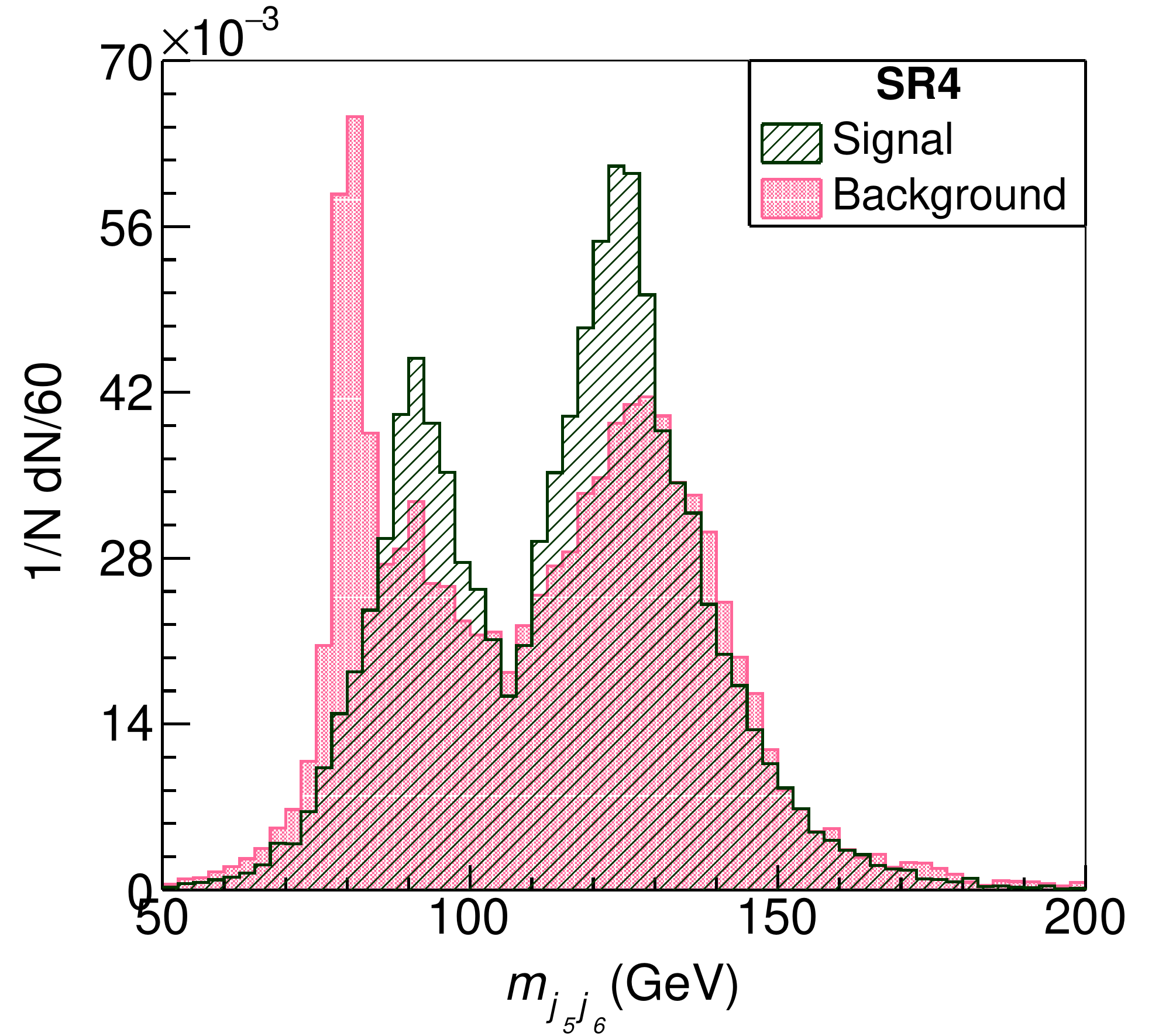}
\includegraphics[width=0.24\columnwidth]{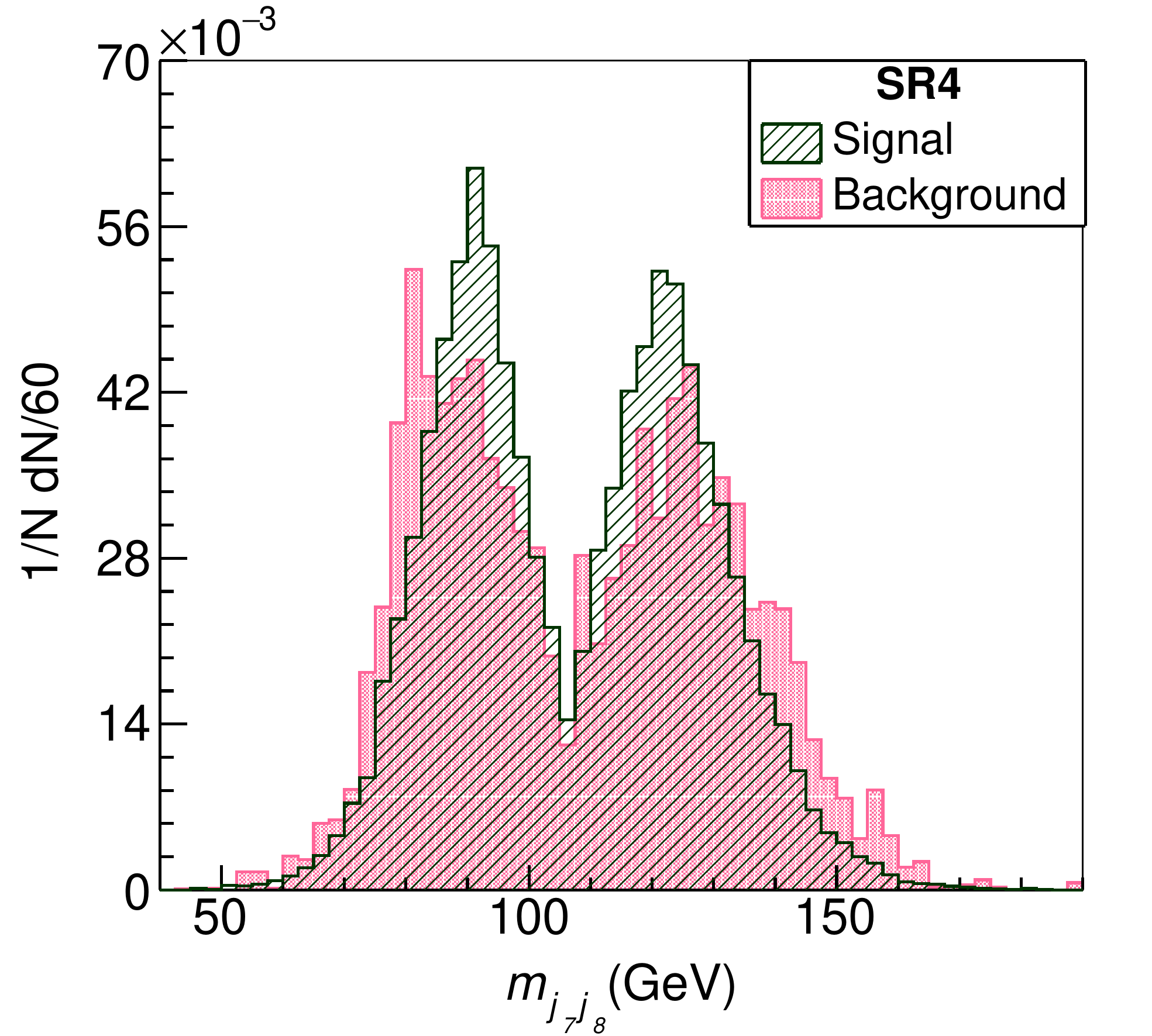}
\caption{Normalised invariant mass distributions for the {\it best-assigned} jet pairs for the signal ({\it BP5}) and background at the 1 TeV $e^-e^+$ collider.}
\label{fig:SR4}
\end{figure}

\vspace{-3mm}
In this SR, $H^0$ and $A^0$ decay dominantly into $hh/ZZ$ and $Zh$, respectively. Thus, we have $hhhZ$ or $ZZZh$ in the final state, in addition to the off-shell $W$-bosons coming from the cascade decays of $H^{\pm \pm}$ and $H^\pm$. The hadronic decays of the $Z/h$-bosons results in up to eight jets in addition to the soft leptons/jets coming from the off-shell $W$-bosons. We require at least seven jets in the final state, with at least two of them to be $b$-tagged. This requirement reduces the background contributions from the diboson and QCD jets (in particular, $b\bar{b}$+jets) production processes to a negligible level, while keeping only a small fraction of contributions from the triboson and multi-top production processes. Brushing aside the soft jets/leptons from the off-shell $W$-bosons, there are as many as eight hard jets in the final state, in addition to the others coming from radiations and pileup interactions. On account of the high-multiplicity signature in the final state, it is particularly burdensome to correctly associate the reconstructed jets in an event with the elementary quarks of that event topology. This makes the kinematic reconstruction of $H^0/A^0$ very challenging. To resolve this {\it combinatorial problem}, we use the so-called $\chi^2$-minimisation method, {\it i.e.} by enumerating and evaluating all possible permutations in an event, we identify the {\it best} assignment through minimising the objective function
\begin{eqnarray*}
\chi^2 &=& \frac{(m_{j_1j_2}-m_{h/Z})^2}{m_{h/Z}^2} + \frac{(m_{j_3j_4}-m_{h/Z})^2}{m_{h/Z}^2} + \frac{(m_{j_5j_6}-m_{h/Z})^2}{m_{h/Z}^2} \\
&+& \frac{(m_{j_7j_8}-m_{h/Z})^2}{m_{h/Z}^2} + \frac{(m_{j_1j_2j_3j_4}-m_{j_5j_6j_7j_8})^2}{\sigma_{H/A}^2},
\end{eqnarray*}
and
\begin{eqnarray*}
\chi^2 &=& \frac{(m_{j_1j_2}-m_{h/Z})^2}{m_{h/Z}^2} + \frac{(m_{j_3j_4}-m_{h/Z})^2}{m_{h/Z}^2} + \frac{(m_{j_5j_6}-m_{h/Z})^2}{m_{h/Z}^2} \\
&+& \frac{(m_{j_1j_2j_3j_4}-m_{j_5j_6j_7})^2}{\sigma_{H/A}^{\prime 2}},
\end{eqnarray*}
respectively, for eight and seven jet events with the jets being denoted as $j_1,j_2,\cdots,j_8$, $\sigma_{H/A} = (m_{j_1j_2j_3j_4}+m_{j_5j_6j_7j_8})/2$ and $\sigma^\prime_{H/A} = (m_{j_1j_2j_3j_4}+m_{j_5j_6j_7})/2$. This method requires enumeration and evaluation of 315 (360) distinct permutations for eight (seven) jet events. In Fig.~\ref{fig:SR4}, we display the normalised invariant mass distributions for the {\it best-assigned}\,\footnote{The jet pairs' assignment corresponding to the minimum objective function is referred to as the {\it best} assignment.} jet pairs for the signal and background events at the 1 TeV $e^-e^+$ collider. The signal events are shown for a benchmark point {\it BP5}: $m_{H^{\pm \pm}}=375$ GeV, $\Delta m=30$ GeV and $v_t \sim 3 \times 10^{-4}$ GeV. The effectiveness of the $\chi^2$-minimisation method has been exemplified by these distributions, each boasting two peaks---one at $m_Z$ and the other at $m_h$, thereby reasonably reconstructing the $h$- and $Z$-bosons in the signal final state. However, note that not only does this method cost a considerable time, but also it obscures the kinematic reconstruction owing to the incorrect assignments of the reconstructed jets to the particles in the event topology.

\newpage

\onecolumngrid

\twocolumngrid

\begin{figure}[htb!]
\centering
\includegraphics[width=0.7\columnwidth]{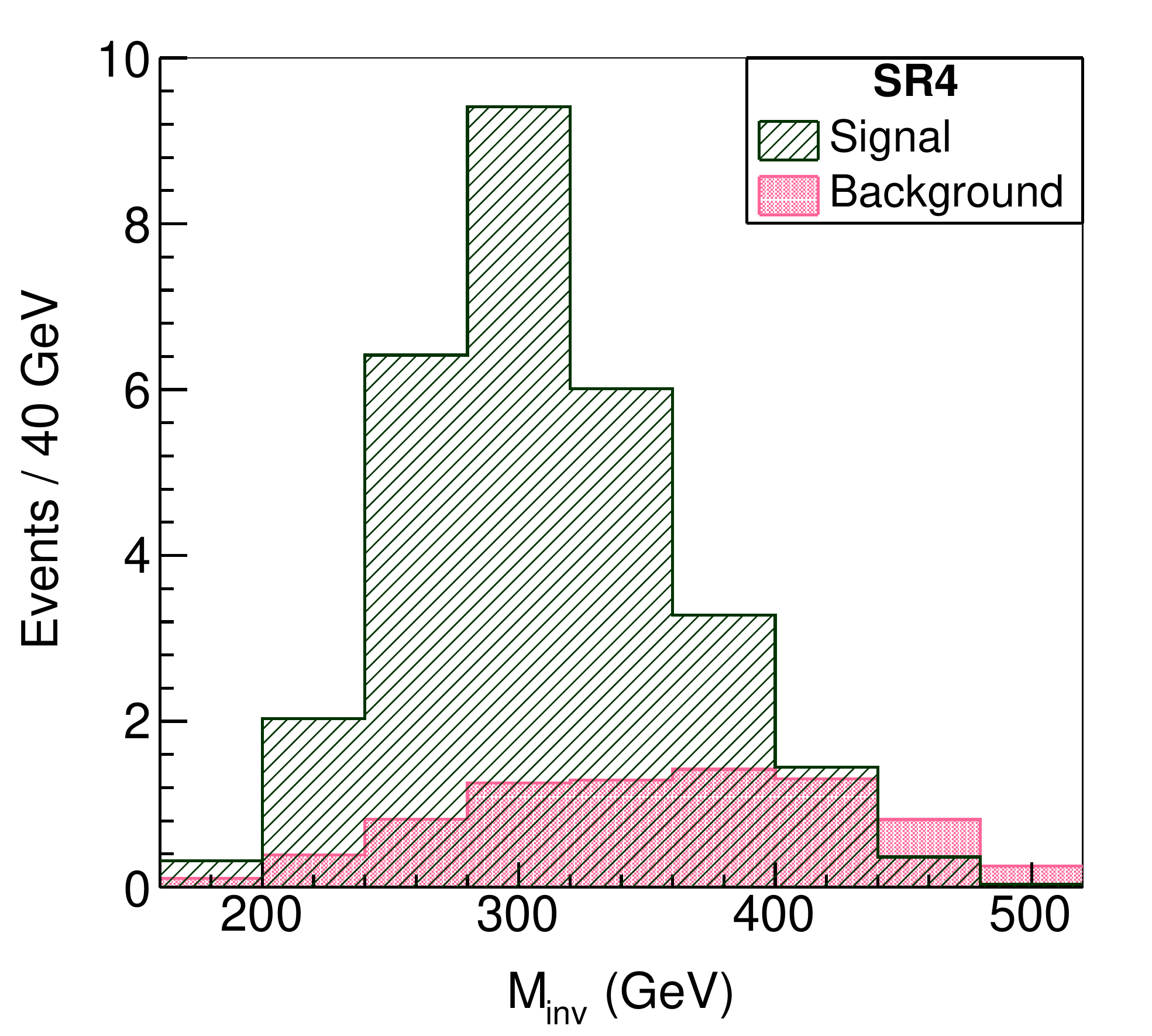}
\caption{$M_{\rm inv}$ distribution for the signal ({\it BP5}) and background. The events are weighted for 1 fb$^{-1}$ luminosity at the 1 TeV $e^-e^+$ collider.}
\label{fig:SR41}
\end{figure}

\noindent Finally, $H^0/A^0$ can be kinematically reconstructed from two pairs of jets in an event:
\[
M_{\rm inv} = (m_{j_1j_2j_3j_4}+m_{j_5j_6j_7j_8})/2,
\]
and
\[
M_{\rm inv} = m_{j_1j_2j_3j_4},
\]
respectively, for 8 and 7 jet events. The sensitivity of this search is increased further by distributing the selected events into 9 bins in the range [160,520] GeV using the invarinat mass $M_{\rm inv}$ (see Fig.~\ref{fig:SR41}). As we expected, the $M_{\rm inv}$ distribution peaks in the 280--320 GeV bin, thereby reconstructing the $H^0/A^0$.

\vspace{-4mm} \subsection{\!\!\!Discovery reach} 
\begin{figure}[]
\centering
\includegraphics[width=0.9\columnwidth]{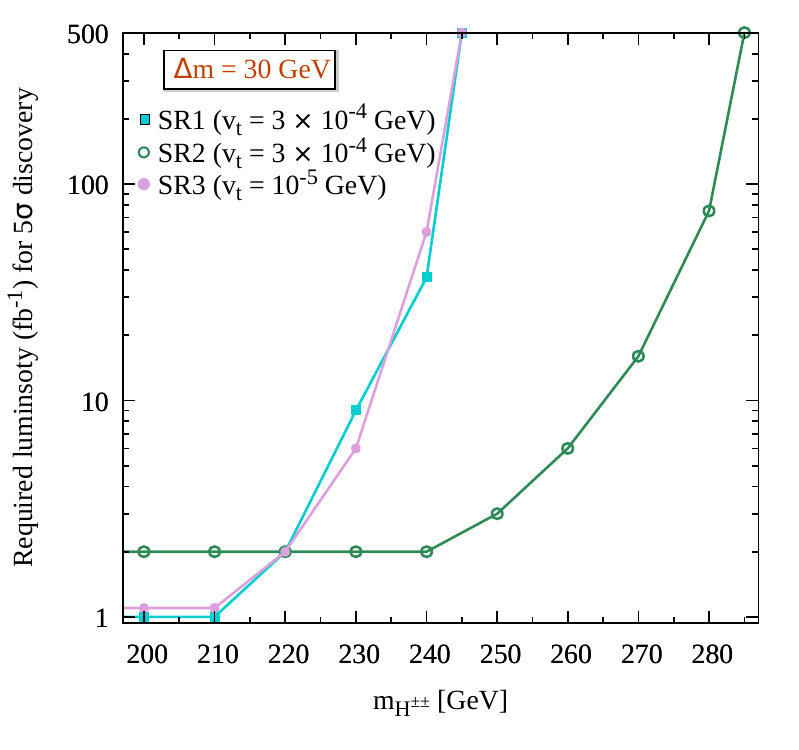}
\includegraphics[width=0.9\columnwidth]{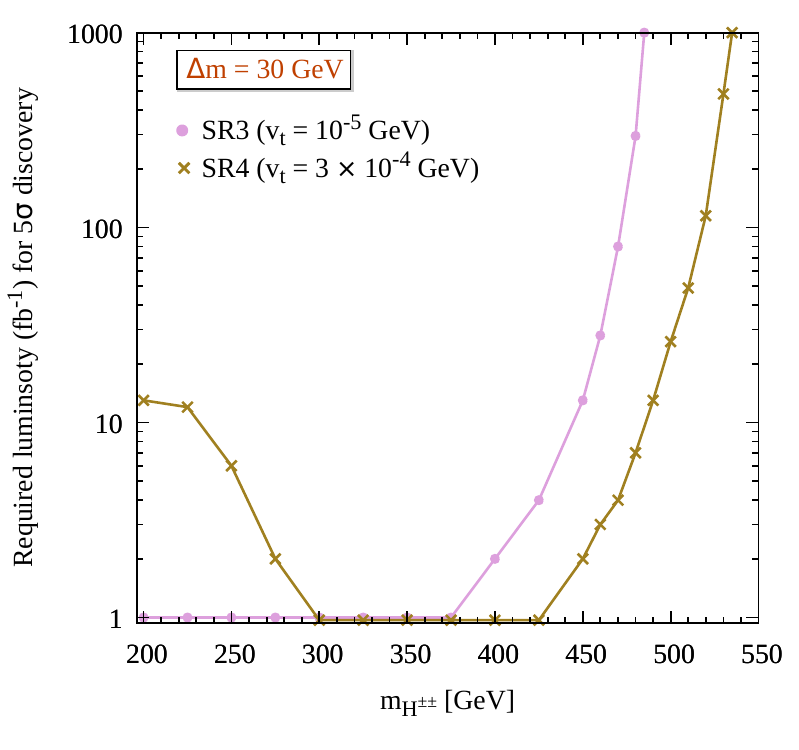}
\caption{Required luminosity for $5\sigma$ discovery of the triplet-like Higgses in different SRs at the 500 GeV (top) and 1 TeV (bottom) $e^-e^+$ colliders.}
\label{fig:lumi}
\end{figure}

\begin{figure}[]
\centering
\includegraphics[width=0.9\columnwidth]{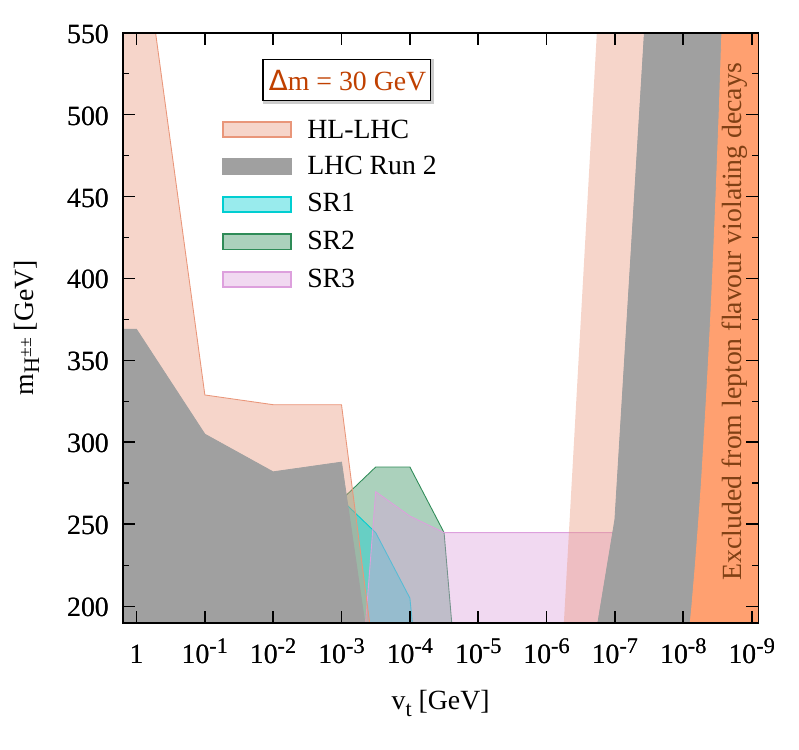}
\includegraphics[width=0.9\columnwidth]{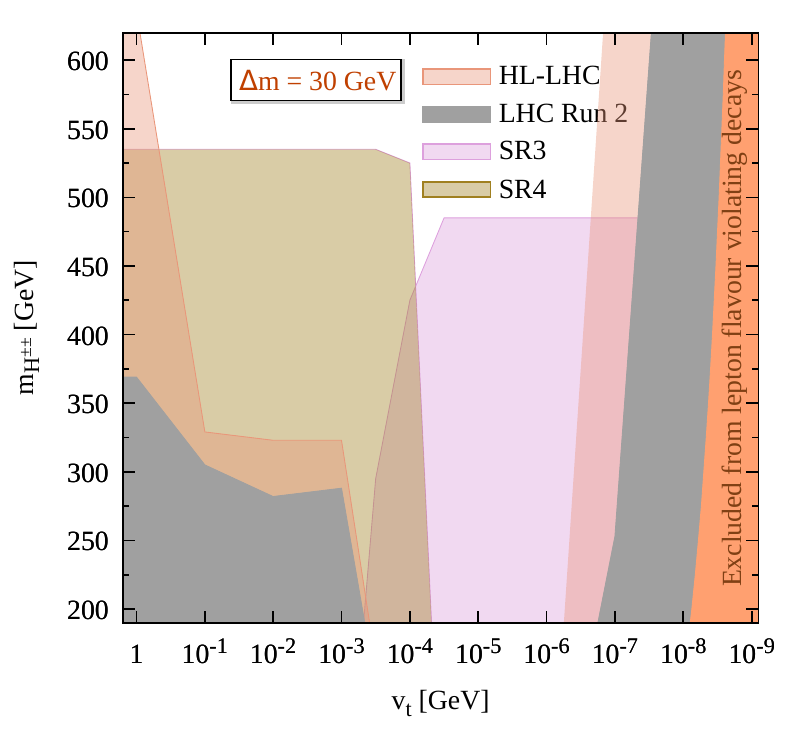}
\caption{Summary of the discovery reaches for the triplet-like Higgses in different SRs at the 500 GeV (top) and 1 TeV (bottom) $e^-e^+$ colliders with 500 and 1000 fb$^{-1}$ luminosity data, respectively. The dark gray and peach shaded regions are taken from Ref.~\cite{Ashanujjaman:2021txz}. See text for details.}
\label{fig:disco}
\end{figure}

\vspace{-3mm}
Next, we estimate the discovery reaches of the above searches by using a hypothesis tester named \texttt{Profile Likelihood Number Counting Combination}, which uses a library of \texttt{C++} classes \texttt{RooFit} \cite{Verkerke:2003ir} in the \texttt{ROOT} \cite{Brun:1997pa} environment. Without going into the intricacy of estimating the background uncertainties, we assume an overall 5\% total uncertainty on the estimated background. In Fig.~\ref{fig:lumi}, we project the required luminosities for $5\sigma$ discovery of the triplet-like Higgses as a function of their mass in different SRs at the 500 GeV and 1 TeV  $e^-e^+$ colliders. Also shown, in Fig.~\ref{fig:disco}, are the discovery reaches for the triplet-like Higgses in the $v_t$-$m_{H^{\pm \pm}}$ plane in different SRs at two configuratioins of $e^-e^+$ colliders---500 GeV and 1 TeV $e^-e^+$ colliders, respectively, with 500 and 1000 fb$^{-1}$ luminosity data. For the sake of completeness, we also show the regions that are excluded from the LHC run 2 searches at 95\% confidence level (shaded in dark gray) or expected to be probed at the HL-LHC (shaded in peach) \cite{Ashanujjaman:2021txz}. For the 500 GeV $e^-e^+$ configuration with 500 fb$^{-1}$ data, {\it SR2} has the maximum discovery reach of $m_{H^{\pm \pm}}\sim 285$ GeV because of its small background, whereas a considerably large background for both {\it SR1} and {\it SR3} limits its discovery reach to $m_{H^{\pm \pm}}\sim 245$ GeV. Likewise, for the 1 TeV $e^-e^+$ configuration with 1000 fb$^{-1}$ data, {\it SR4} is the most promising signal region with the discovery reach of $m_{H^{\pm \pm}}\sim 535$ GeV because of its small background and ability to kinematically reconstruct the neutral Higgses $H^0/A^0$, and {\it SR3} has a discovery reach of $m_{H^{\pm \pm}}\sim 485$ GeV.

\vspace{-4mm} \section{\label{sec:conclusion}\!\!\!Summary and outlook} \vspace{-3mm}
While the triplet-like Higgses up to a few hundred GeV masses are already excluded for a vast region of the model parameter space from the LHC searches, strikingly, there is a region of this parameter space---with large enough positive mass-splitting between the doubly- and singly-charged Higgses and moderate triplet Higgs scalar vacuum expectation value---that is beyond the reach of the existing LHC searches, and such Higgses as light as 200 GeV or even lighter are still allowed by the LHC data \cite{Ashanujjaman:2021txz}. In this region of the parameter space, the charged Higgses decay exclusively to the neutral ones and off-shell $W$-bosons. The latter results in soft leptons/jets that are challenging to reconstruct at the LHC. Furthermore, the neutral Higgses decay into neutrinos or $b\bar{b},t\bar{t},ZZ,Zh,hh$, thereby resulting in final states that are challenging to probe at the LHC owing to the towering SM backgrounds. However, owing to a cleaner environment, future lepton colliders are expected to have better prospects for probing this LHC elusive parameter space. In this work, we study several search strategies targeting different parts of this parameter space at future $e^-e^+$ colliders with 500 GeV and 1 TeV centre of mass energies. We find that a vast region of this parameter space, with some parts of this---parametrised by {\it SR1} and {\it SR4} signal regions---allowing for kinematic reconstructions of the triplet-like neutral Higgses, could be probed with $5\sigma$ discovery with the early $e^-e^+$ colliders' data. In closing this section, here are a few comments in order. $(i)$ For the sake of definiteness, we have shown our findings for a mass-splitting of 30 GeV. However, the searches presented above would also be sensitive for smaller mass-splittings. Brushing aside a little quantitative difference in production cross-sections for the triplet-like Higgses, the only major difference between a smaller mass-splitting (say, 10 GeV) and a larger one (say, 30 GeV) is that the leptons/jets stemming from the off-shell $W$-bosons would be softer. Note that only the search in SR3 targets soft leptons/jets in the final state. Therefore, for a smaller mass-splitting, the $E_{\rm eff}$ distribution in Fig.~\ref{fig:SR3} would shift towards lower values, thus, allowing one to impose a stronger cut on it. While this would impinge only a little on the signal strength, the same would significantly enhance the signal-to-background ratio, primarily on account of a much reduced background. On the other hand, the searches in SR1, SR2 and SR4 targets hard jets/leptons stemming from the triplet-like Higgses' decays, thus these are almost independent of the mass-splitting. $(ii)$ The SR3 region of parameter space could also be probed by using a search with a photon and missing transverse momentum in the final state. Such a final states suffer from a large irreducible background contributions from the $t$-channel $W$-exchange process $\nu\bar{\nu}\gamma$. Though this background could be reduced by a factor of few by using polarised beams (positive for $e^-$ and negative for $e^+$), on account of the small signal strength occasioned by the requirement of an energetic photon, the discovery reach of such a search would be very limited.

\vspace{-4mm} \acknowledgments \vspace{-3mm} KG acknowledges support from the DST INSPIRE Research Grant [DST/INSPIRE/04/2014/002158] and SERB Core Research Grant [CRG/2019/006831].

\newpage

\bibliography{v0}

\end{document}